\documentclass[12pt,a4paper]{article}

\usepackage{amsmath}
\usepackage{epsf}
\usepackage{amsthm,amssymb}

\allowdisplaybreaks

\renewcommand{\baselinestretch}{1.5}

\newcommand{\be}{\begin{equation}} 
\newcommand{\ee}{\end{equation}} 
\newcommand{\bea}{\begin{eqnarray}}  
\newcommand{\eea}{\end{eqnarray}} 
 
\newcommand{\mcM}{\mathcal{M}}
\newcommand{\cO}{\mathcal{O}}
\newcommand{\cR}{\mathcal{R}}
\newcommand{\ad}{\dot{a}}
\newcommand{\add}{\ddot{a}}
\newcommand{\rd}{\dot{\rho}}
\newcommand{\Gd}{\dot{G}}
\newcommand{\Ld}{\dot{\Lambda}}
\newcommand{\ap}{a^{\prime}}
\newcommand{\Lp}{\Lambda^{\prime}}
\newcommand{\Gp}{G^{\prime}}

\newcommand{\bt}{\tilde{\beta}}
\makeatletter \@addtoreset{equation}{section} \makeatother
\renewcommand{\theequation}{\thesection.\arabic{equation}}
\begin{document} 
\begin{titlepage}
\begin{center}
\renewcommand{\baselinestretch}{1} \large\normalsize
\hfill MZ-TH/05-14\\
\hfill ITP-UU-05/27 \\
\hfill SPIN-05/21 \\
\renewcommand{\baselinestretch}{1.5} \large\normalsize
\vskip 0.9cm {\large \bf  
From Big Bang to Asymptotic de Sitter: \\
Complete Cosmologies in a Quantum Gravity Framework}

\vskip 0.9cm

{\bf M.\ Reuter$^a$ and F.\ Saueressig$^b$}  \\[3ex]
\renewcommand{\baselinestretch}{1} \large\normalsize

$^a${\em Institute of Physics, University of Mainz,\\[1ex] 
 Staudingerweg 7, D-55099 Mainz, Germany} \\[2ex]
$^b${\em Institute of Theoretical Physics} \& {\em Spinoza Institute, \\[1ex]
Utrecht University, 3508 TD Utrecht, The Netherlands} \\ 
\renewcommand{\baselinestretch}{1.5} \large\normalsize
\end{center}
\vskip 0.9cm
\begin{center} {\bf ABSTRACT } \end{center}
\vspace{-2mm}
Using the Einstein-Hilbert approximation of asymptotically safe quantum gravity we present a consistent renormalization group based framework for the inclusion of quantum gravitational effects 
into the cosmological field equations. Relating the renormalization group scale to cosmological time via a dynamical cutoff identification this framework applies to all stages of the cosmological evolution. The very early universe is found to contain a period of ``oscillatory inflation'' with an infinite sequence of time intervals during which the expansion alternates between acceleration and deceleration. For asymptotically late times we identify a mechanism which prevents the universe from leaving the domain of validity of the Einstein-Hilbert approximation and obtain a classical de Sitter era.
\noindent 
\end{titlepage}
\section{Introduction}
It is generally believed that in the very early universe, at times smaller than the Planck time, quantum gravity effects play a crucial role. Since the physics in the Planck era prepares the initial conditions for the subsequent classical evolution of the universe and, as a result, is likely to be responsible for many of its features we observe today it is clearly very desirable  to gain some understanding of the quantum gravitational processes which took place immediately after the Big Bang.

The construction of a consistent quantum theory of gravity which is indispensable for investigations of this kind is a major challenge, of course. In trying to set up such a theory the first question to ask is what are the degrees of freedom which properly describe the gravitational field at the quantum level. In view of the successes of classical general relativity (GR) the most natural working hypothesis is that they are given by the gauge invariant content of the metric tensor $g_{\mu \nu}$. Then, trying to construct a quantum field theory based upon $g_{\mu \nu}$, a second question arises: what is the (bare) action describing its dynamics? The obvious first option to try is the Einstein-Hilbert action underlying classical general relativity, but it is well-known that its perturbative quantization leads to a non-renormalizable theory. The next logical possibility is to stay within perturbation theory but to use a different action. For example, one could try to add terms quadratic in the curvature to the Einstein-Hilbert term; this yields a theory which is indeed perturbatively renormalizable but nevertheless has to be discarded because of unitarity problems. 

Thus it seems likely that a satisfactory quantum field theory of the continuum metric, if it exists, requires us to abandon both the classical action and perturbation theory. Weinberg's ``asymptotic safety'' scenario \cite{wein} is an approach in precisely this category\footnote{Approaches which (typically) retain the Einstein-Hilbert form of the action but instead abandon the traditional continuum $g_{\mu \nu}$ include loop gravity \cite{ash} and random triangulations \cite{loll}.}. Here the basic idea is to define, non-perturbatively, a theory which is based upon a very special bare action, namely one which is infinitesimally close to a fixed point $\Gamma_{*}[g_{\mu \nu}]$ of the Wilsonian renormalization group (RG). If it exists, the fixed point allows us to take the limit of an infinite ultraviolet (UV) cutoff in a controlled, or ``safe'' way. 

The theory thus constructed will be referred to as Quantum Einstein Gravity or ``QEG''. Using the technique of the exact RG equations in the continuum \cite{bagber} - \cite{ym} a considerable amount of evidence has been collected by now \cite{mr}-\cite{proper} which indicates that an appropriate fixed point is indeed likely to exist on the ``theory space'' consisting of the diffeomorphism invariant functionals of $g_{\mu \nu}$. Practical calculations typically involve a truncation of this theory space. A conceptually quite different approximation leading to similar conclusions is the (exact) quantization of the restricted class of metrics admitting two Killing vectors \cite{max}.
\begin{figure}[t]
\renewcommand{\baselinestretch}{1}
\epsfxsize=0.8\textwidth
\begin{center}
\leavevmode
\epsffile{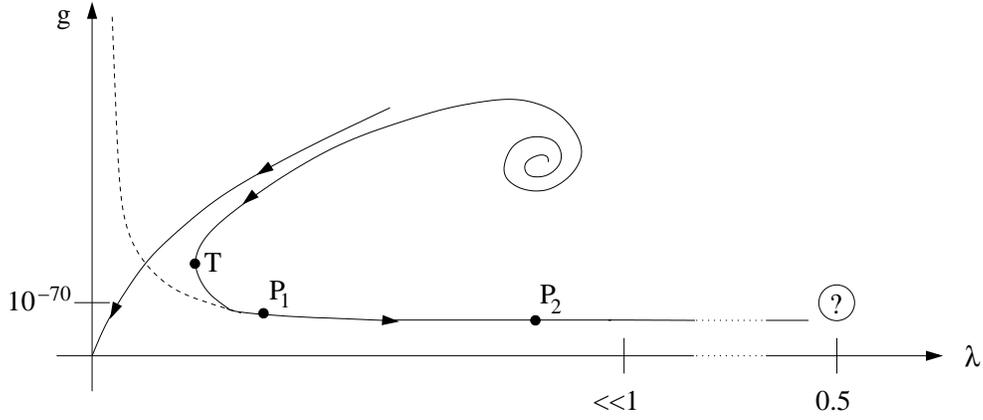}
\end{center}
\parbox[c]{\textwidth}{\caption{\label{null}{\footnotesize The Type IIIa trajectory realized in Nature and the separatrix. The dashed line is a trajectory of the canonical RG flow. (Taken from ref.\ \cite{h3}.)}}} 
\end{figure}

Most of the investigations using the effective average action formalism \cite{avact} - \cite{ym} for implementing the RG ``coarse graining'' employ the so-called Einstein-Hilbert truncation which retains only Newton's constant $G(k)$ and the cosmological constant $\Lambda(k)$ as running parameters. Here $k$ denotes the variable infrared (IR) cutoff, the momentum scale down to which the quantum fluctuations of the metric are integrated out. If one introduces the dimensionless couplings $g(k) \equiv k^2 G(k)$ and $\lambda(k) \equiv \Lambda(k)/k^2$ the flow equations in this truncation consist of an autonomous system of two coupled ordinary differential equations: 
\be\label{1.auto}
k \partial_k g = \beta_g(g, \lambda) \; , \qquad  k \partial_k \lambda = \beta_\lambda(g, \lambda). 
\ee
The beta-functions $\beta_g$ and $\beta_\lambda$ were first derived in \cite{mr}, and the resulting equations were studied numerically in ref.\ \cite{frank1}. The flow on the $g$-$\lambda$-plane displays two fixed points: a Gaussian fixed point (GFP) at the origin, and a non-Gaussian fixed point (NGFP) at $(g_{*}, \lambda_{*})$ with $g_{*}$ and $\lambda_{*}$ both positive. At least according to what can be said on the basis of the Einstein-Hilbert truncation, this NGFP has all the properties necessary for asymptotic safety. 

Ref.\ \cite{frank1} contains a complete classification of all types of RG trajectories occurring in the Einstein-Hilbert truncation. Particularly important are those of Type Ia, IIa, and IIIa which, when $k$ is lowered, run towards negative, vanishing, and positive values of the cosmological constant, respectively. In ref.\ \cite{h3} the very special trajectory which is realized in Nature has been identified and its parameters were determined. This trajectory is of Type IIIa; it is sketched schematically in fig. \ref{null}. 

For $k \rightarrow \infty$ it starts infinitesimally close to the NGFP. Then, lowering $k$, the trajectory spirals about the NGFP and approaches the ``separatrix'', the distinguished trajectory which ends at the GFP. It runs almost parallel to the separatrix for a very long ``RG time''; only in the ``very last moment'' before reaching the GFP, at the turning point T, it gets driven away towards larger values of $\lambda$. In fig.\ \ref{null} the points P$_1$ and P$_2$ symbolize the beginning and the end of the regime in which classical general relativity is valid (``GR regime''). This section of the trajectory is virtually identical to a canonical one for which $G$ and $\Lambda$ are $k$-independent. The classical regime starts soon after the turning point T which is passed at the scale $k_{\rm T} \approx 10^{-30} m_{\rm Pl}$ corresponding to the (macroscopic!) length $k_{\rm T}^{-1} \approx 10^{-3}$ cm.\footnote{Since it is only the cosmological constant which shows a significant running between $k=m_{\rm Pl}$ and $k = k_{\rm T}$, and since $\Lambda$  cannot probably be measured in a millimeter-size laboratory experiment, one nevertheless would not expect to see violations of classical general relativity immediately above $k_{\rm T}$ \cite{h3}.} 

In ref.\ \cite{h3} it was argued that  there starts a regime of strong IR renormalization effects to the right of the point P$_2$ which might become visible at astrophysical and cosmological length scales. In fact, within the Einstein-Hilbert approximation, trajectories of Type IIIa cannot be continued to the extreme IR ($k \rightarrow 0$). They terminate at a non-zero value of $k$ as soon as the trajectory reaches $\lambda = 1/2$. (Close to the question mark in fig.\ \ref{null}.) Before it starts becoming invalid and has to be replaced by a more precise treatment, the Einstein-Hilbert approximation suggests that $G$ will increase, while $\Lambda$ decreases, as $\lambda \nearrow  1/2$ \cite{frank1}.

The Type IIIa trajectory of QEG which Nature has selected is highly special or ``unnatural'' in the following sense. It is fine-tuned in such a way that it gets {\it extremely} close to the GFP before ``turning left''. In \cite{h3} it was shown that the coordinates $g_{\rm T}$ and $\lambda_{\rm T}$ of the turning point are both very small: $g_{\rm T} = \lambda_{\rm T} \approx 10^{-60}$. In the GR regime, $g$ decreases from $g(k) = 10^{-70}$ at a typical terrestrial length scale of $k^{-1} = 1$ meter to $g(k) = 10^{-92}$ at the solar system scale of $k^{-1} = 1$ astronomical unit. Extrapolating to cosmological scales one finally has $g(k) = 10^{-120}$ when $k$ equals the present Hubble constant $H_0$. 

In this analysis the two free parameters which uniquely characterize a Type IIIa trajectory where derived from the measured values of $G(k)$ for laboratory values of $k$, and $\Lambda(k)$ at $k \approx H_0$. It can be argued that the IR renormalization effects, if they exist, could not change $G$ and $\Lambda$ by many orders of magnitude between $P_2$ (at solar system scales, say) and cosmological scales.

Indeed, the present Hubble parameter $k = H_0$ is approximately the scale where the Einstein-Hilbert trajectory becomes unreliable. The observations indicate that today the cosmological constant is of the order $H_0^2$. Interpreting this value as the running $\Lambda(k)$ at the scale $k = H_0$, the dimensionless $\lambda(k)$, at this scale, is of the order unity: $\lambda(H_0) \equiv \Lambda(H_0)/H_0^2 = \cO(1)$. So it is quite precisely near the present Hubble scale where the Einstein-Hilbert truncation becomes insufficient for a description of the trajectory Nature has chosen. 

The ``unnaturalness'' of Nature's gravitational RG trajectory has an important consequence. Because it gets so extremely close to the GFP it spends a very long RG time in its vicinity because the $\beta$-functions are small there. As a result, the termination of the trajectory at $\lambda = 1/2$ is extremely delayed, by 60 orders of magnitude, compared to a generic trajectory where this happens for $k$ near the Planck mass. In ref.\ \cite{h3} it was argued that this non-generic feature of the trajectory is a necessary condition for a long classical regime with $G, \Lambda = const$ to emerge, and any form of classical physics to be applicable. 

It was also shown that for any trajectory which actually does admit a long classical regime the cosmological constant is automatically small. Nevertheless, the fine-tuning behind the ``unnatural'' trajectory Nature has selected is of a much more general kind than the traditional cosmological constant problem \cite{cc1,cc2,cc3,cc4}: the primary issue is the emergence of a classical space-time; once this is achieved, the extreme smallness of the observed $\Lambda$ (compared to $m_{\rm Pl}^2$) comes for free. (See \cite{h3} for a detailed discussion.)

Knowing at least the qualitative features of the RG running of the gravitational parameters one can try to use this information for investigating how quantum gravity effects modify the classical Friedmann-Robertson-Walker (FRW) cosmology. An immediate consequence of an approach of this kind is that the cosmological constant becomes a time-dependent quantity \cite{mrcw} so that in principle it should be possible to understand its (tiny) value today as the result of a dynamical evolution process. In this respect the RG approach has certain features in common with the quintessence models \cite{cwquint,quint}. The main difference is that the evolution is driven by the vacuum fluctuations of the gravitational field itself, and no extra quintessence field needs to be introduced.

In ref.\ \cite{cosmo1} a first application of the RG flow from QEG to the cosmology of the early universe has been described. The tool used there was a kind of gravitational ``RG improvement'' \cite{nelson} - \cite{set}. The running couplings $G(k)$ and $\Lambda(k)$ where converted to functions of the cosmological time $t$ by identifying the RG scale $k$ with $1/t$ or, what was the same in this context, the Hubble parameter $H(t)$. Then the resulting $G(t)$ and $\Lambda(t)$ were inserted into the Einstein equations for a homogeneous and isotropic universe. In the fixed point regime where    
\be\label{1.1}
G(k) = g_{*} / k^2 \; , \quad \Lambda(k) = \lambda_{*} \, k^2 \, ,
\ee
and $G(t) \propto t^2$, $\Lambda(t) \propto 1/t^2$ it was then possible to find exact analytic solutions for the Robertson-Walker scale factor $a(t)$. Provided this improvement of the cosmological evolution equations does indeed encapsulate the leading quantum effects, the fixed point solution should describe the universe at times much earlier than the Planck time since \eqref{1.1} is valid only as long as $k \gg m_{\rm Pl}$, the Planck mass being the lower boundary of the asymptotic scaling region. A strong argument in favor of the validity of the approach is that gravity becomes weakly coupled for $t \rightarrow 0$ since $G(k)$, vanishing for $k \rightarrow \infty$, is an asymptotically free coupling.

In particular for the equation of state $p = \rho/3$ and flat 3-sections ($K=0$) the fixed point cosmology has various remarkable properties.\footnote{In ref.\ \cite{cosmo1} also fixed point solutions for the more general equation of state $p = w \rho$ and (pseudo)spherical 3-spaces ($K = \pm 1$)  were obtained. In a different context similar cosmologies were investigated in \cite{kall}. Fixed point cosmologies of the late universe are discussed in \cite{cosmo2,elo}.} It is completely scale-free; the scale factor and the matter energy density behave as $a(t) \propto t$ and $\rho(t) \propto 1/t^4$, respectively. For every RG trajectory interpolating between the fixed point behavior \eqref{1.1} and the classical $G,\Lambda =  const$, the fixed point solution is a {\it universal attractor} in the space of improved FRW cosmologies in the sense that they all approach the fixed point solution for $t \rightarrow 0$. Stated the other way around: the fixed point cosmology prepares the initial conditions for the subsequent cosmological evolution which is essentially classical. Those initial conditions are such that the energy density today equals precisely the critical one, $\rho_{\rm crit}$. No fine-tuning is necessary here, only a discrete choice, picking flat rather than (pseudo)spherical 3-spaces 
\cite{cosmo1}. Furthermore, as a result of the linear expansion $a(t) \propto t$ at early times, the RG improved cosmology has no particle horizon. This is another feature which could be of phenomenological importance.

Up to now the solution to the coupled system of RG- and cosmological evolution equations is known only in the fixed point regime. The purpose of the present paper is to analyze the complete solution all the way from the Big Bang to scales of the order of the present Hubble constant. Using mostly numerical methods we shall discuss in detail whether or not the Einstein-Hilbert truncation is sufficient in order to connect the UV fixed point regime (Planck era) to the present era of the universe.

The remaining sections of this paper are organized as follows. In Section 2 we briefly review some properties of the coupled RG- and cosmological evolution equation, point out a conceptual difficulty related to the fact that this system is overdetermined for any ``rigid'' cutoff identification, and propose a solution to this problem. Then, in Section 3, we present the RG equations to be used in the following. Section 4 contains an investigation of all those properties of the RG improved cosmology which do not depend on the detailed form of the RG trajectory but rather reflect certain general properties of the ``theory space'' on which the RG flow takes place. In Section 5 we explore the actual cosmological time evolution; in doing so we advocate the point of view that this evolution is primarily an evolution with respect to the scale $k$. Section 6 contains a brief summary and discussion of the results.

\section{RG improved Einstein's equation}
\label{sect2}
Following \cite{cosmo1,cosmo2} we investigate homogeneous and isotropic cosmologies described by a standard Robertson-Walker metric containing the scale factor $a(t)$ and the parameter $K = 0, \pm 1$ which distinguishes the three possible types of maximally symmetric 3-spaces of constant cosmological time $t$. The dynamics is governed by Einstein's equation $G_{\mu \nu} = - \Lambda g_{\mu \nu} + 8 \pi G T_{\mu \nu}$ where $T_\mu^{~\nu} \equiv {\rm diag}[-\rho, p, p, p]$ is a conserved energy momentum tensor for which the equation of state $p(t) = w \rho(t)$ with an arbitrary constant $w > -1$  is assumed.

To start with, we assume that we have fixed a certain ``cutoff identification'' $k = k(t)$. Then every RG trajectory $k \mapsto \left( G(k), \Lambda(k) \right)$ obtained by solving the RG equations gives rise to time dependent functions $G(t) \equiv G(k=k(t))$ and $\Lambda(t) \equiv \Lambda(k = k(t))$. Replacing the constants $G$ and $\Lambda$ in Einstein's equation with these functions, and specializing for a Robertson-Walker metric, the field equations boil down to
\begin{align*}
\addtocounter{equation}{1} 
\tag{\theequation a} 
\label{2.1a} 
\left( \frac{\ad}{a} \right)^2 + \frac{K}{a^2}  = \frac{1}{3} \, \Lambda + \frac{8 \pi}{3} \, G \, \rho \,  , \\  
\tag{\theequation b}
\label{2.1b} 
\rd + 3 \, (1+w) \, \frac{\ad}{a} \, \rho  = 0 \, ,\\ 
\tag{\theequation c} 
\label{2.1c} 
\Ld + 8 \, \pi \, \rho \, \Gd  = 0 \, . 
\end{align*} 
Here the ``dot'' denotes the derivative with respect to the cosmological time $t$. Eq.\ \eqref{2.1a} is the familiar Friedmann equation, albeit with a time dependent $G$ and 
$\Lambda$, and eq. \eqref{2.1b} is the conservation law $D_{\mu} T^\mu_{~\nu} = 0$. Eq.\ \eqref{2.1c} represents an additional consistency condition which is necessary for the integrability of Einstein's equation. If \eqref{2.1c} holds true, its RHS has a vanishing covariant divergence, as has its LHS by virtue of Bianchi's identity.

For $G(t)$ and $\Lambda(t)$ fixed, the equations (\ref{2.1a},b,c) constitute a system of 3 equations for only two unknowns, $a(t)$ and $\rho(t)$. In general this system is overdetermined and, for generic $G(t)$ and $\Lambda(t)$, will not admit any solution for $a(t)$ and $\rho(t)$. This can be seen as follows. The time dependence of $\rho$ follows directly from \eqref{2.1c},
\be\label{2.2}
\rho(t) = - \, \frac{1}{8 \pi} \, \frac{\Ld(t)}{\Gd(t)} \, , 
\ee
and integrating \eqref{2.1b} yields $\rho a^{3+3w} = \mcM / (8 \pi)$, with $\mcM$ a constant of integration. Combining this relation with \eqref{2.2} we find the time dependence of the scale factor:
\be\label{2.3}
a(t) = \left[ - \, \mcM \, \frac{\Gd(t)}{\Ld(t)}\right]^{1/(3+3w)} \, .
\ee
Already at this point we know $a(t)$ and $\rho(t)$, but we have not used the improved Friedmann equation \eqref{2.1a} yet. Checking whether \eqref{2.2} and \eqref{2.3} also solve \eqref{2.1a}, one usually discovers that this is not the case: the three equations (\ref{2.1a},b,c) are not integrable for generic $G(t)$ and $\Lambda(t)$. 

At this  point there are two logical possibilities: either the improvement of the field equations is not an adequate way of exploiting the RG information, or one can ultimately arrive at a description which is self-consistent within this framework by modifying $G(t)$ and $\Lambda(t)$ in such a way that all three equations (\ref{2.1a},b,c) are satisfied. As for the second possibility, there are two obvious ways of modifying the time dependence of $G$ and $\Lambda$. On can change the cutoff scheme \cite{mr,cosmo1}, i.e., the operator $\cR_k(-D^2)$ which suppresses the ``slow'' modes in the path integral, thus slightly changing the trajectory $k \mapsto (G(k), \Lambda(k))$, or one can change the ``cutoff identification'' $k = k(t)$. The details of both the trajectory\footnote{The critical exponents of the fixed points, for instance, are universal (i.e., $\cR_k$-independent) features of the RG flow, but not the precise shape of the trajectories.} and the cutoff identification are non-universal, i.e., unphysical, but in combining $G(k)$ with $k = k(t)$ to obtain $G(t) \equiv G(k = k(t))$ the non-universalities can cancel to some extent,\footnote{See ref.\ \cite{bh} for a particularly transparent example in a black hole context.} provided $k(t)$ is chosen appropriately. 

The length scale set by the IR cutoff\footnote{See ref.\ \cite{oliverfrac} for a more detailed discussion.}, $\ell \approx k^{-1}$, can be visualized as the variable resolution of the ``microscope'' with which space-time is observed \cite{avactrev} since the effective average action $\Gamma_k$ describes the dynamics of fields averaged over volumes of linear extension $\ell$ \cite{avact}. In cosmological applications it is sensible to readjust the resolution of the microscope as the universe becomes larger. In which way precisely $\ell$ is increased as the scale factor grows depends on the interpretation one wants to give to the coarse grained picture of the expanding universe. In \cite{cosmo1} the ansatz $k \propto 1/t$ was used, motivated by the fact that when the age of the universe is $t$, no process with frequency smaller than $1/t$ can have occurred yet. Also the Hubble scale $k \propto H(t)$ would be a natural choice since in cosmology the Hubble length $\ell_{H} \equiv 1/H(t)$ measures the size of the ``Einstein elevator'' outside which curvature effects become appreciable.

In the present paper we shall adopt the following strategy for choosing $k(t)$. In order to circumvent the problem of the overdetermined equations (\ref{2.1a},b,c) we shall not a priori fix the cutoff identification $k = k(t)$ in a rigid way, but rather {\it derive} it from the field equations themselves, imposing the condition that those equations should be integrable. On the RG side, we shall stick to a fixed cutoff scheme. Then, choosing initial conditions for the trajectory, we obtain uniquely defined functions $G(k)$ and $\Lambda(k)$. Next it is important to note that, given the $k$-dependence of $G$ and $\Lambda$, we immediately know $a$ and $\rho$ as functions of $k$:
\begin{align*}
\addtocounter{equation}{1} 
\tag{\theequation a} 
\label{2.4a} 
a(t(k)) & = \left[ - \mcM \, \frac{ G^{\prime}(k)}{ \Lambda^{\prime}(k) } \right]^{1/(3+3w)} \, , \\
\tag{\theequation b}
\label{2.4b} 
\rho(t(k)) & = - \, \frac{1}{8 \pi} \, \frac{\Lambda^{\prime}(k) }{G^{\prime}(k)} \, .
\end{align*} 
On the LHS of these relations $t(k)$ is the functional inverse of the cutoff identification $k = k(t)$. The eqs.\ (\ref{2.4a},b) follow from \eqref{2.2} and \eqref{2.3}, respectively, since $\Gd = \left( dk / d t \right) G^{\prime}$, $\Ld = \left(dk / d t \right) \Lambda^{\prime}$, and the derivative of $k(t)$ drops out from the ratio $\Gd / \Ld$. (Here and in the following the prime denotes a derivative with respect to $k$.)

With (\ref{2.4a},b) the second and the third of the eqs.\ (\ref{2.1a},b,c) are satisfied. We can now try to make the whole system consistent by allowing the Friedmann equation \eqref{2.1a} to fix the relationship between $k$ and $t$. Inserting (\ref{2.4a},b) into eq.\ \eqref{2.1a} the latter yields the following differential equation\footnote{Here we have chosen the sign of the square root such that $t$ increases with decreasing $k$, i.e., late cosmological times will be associated with small $k$-values on the RG-trajectory.} for $k = k(t)$ 
\be\label{1.2}
\frac{d \, k}{d \, t} =  \frac{a}{\ap} \, \left[ \frac{\Lambda}{3} + \frac{8 \pi}{3} \, G \, \rho \right]^{1/2} \, , 
\ee
which can be solved (numerically) for any given trajectory.\footnote{A similar strategy was used in ref.\ \cite{set}.} The RHS of \eqref{1.2} is completely specified in terms of RG data: $G(k)$ and $\Lambda(k)$ obtained directly from the trajectory, and $a$ and $\rho$ are given by (\ref{2.4a},b).

This method provides us with a solution $\{ a(t), \rho(t), k(t) \}$ for any trajectory $\left( G(k), \Lambda(k) \right)$, but it is not clear a priori whether this solution is physically sensible. It can have a meaningful interpretation only if the continuous readjustment of the ``microscope's'' resolution described by $k = k(t)$ is correlated with the expansion of the universe in a transparent and, in particular, monotonic way. This will indeed turn out to be the case: We shall see that the function $k(t)$ obtained dynamically is reasonably close to $k(t) \propto H(t)$ during all epochs of the history of the universe.

\section{The RG equations}
\label{sect3}

The RG equations of QEG in the Einstein-Hilbert truncation are given by \eqref{1.auto}. The pertinent theory space is the $g$-$\lambda$-plane coordinatized by the dimensionless couplings $g$ and $\lambda$. The (dimensionless, $k$-independent) beta-functions $\beta_g$ and $\beta_\lambda$ are the components of a vector field on the theory space. In ref.\ \cite{mr} they were obtained for $d$ space-time dimensions.

In $d$ dimensions, the dimensionful couplings are given by $G(k) \equiv k^{2-d} g(k)$ and $\Lambda(k) \equiv k^2 \lambda(k)$. Their $k$-dependence is governed by the system of equations
\be\label{4.1}
k \, \frac{d}{d k} \, G(k) = \beta_G(G, \Lambda, k) \, , \quad k \, \frac{d}{d k} \, \Lambda(k) = \beta_{\Lambda}(G, \Lambda, k) \, . 
\ee
The (dimensionful, $k$-dependent) beta-functions $\beta_G$ and $\beta_\Lambda$ read \cite{mr}:
\begin{align*}
\addtocounter{equation}{1} 
\tag{\theequation a} 
\label{4.2a} 
\beta_G(G, \Lambda, k)  = & \eta_N \, G \, , \\
\tag{\theequation b}
\label{4.2b} 
\beta_{\Lambda}(G, \Lambda, k)  = & \eta_N \, \Lambda + \frac{1}{2} (4 \pi)^{1-d/2} \, k^d \, G \, \bigg[2d(d+1) \Phi^1_{d/2}(-2 \, \Lambda/k^2) \\ & \qquad - 8 d \Phi^1_{d/2}(0) - d(d+1) \eta_N \tilde{\Phi}^1_{d/2}(-2 \, \Lambda/k^2) \bigg] \, .
\end{align*} 
Here $\eta_N$ denotes the anomalous dimension of $\sqrt{g} R$; it can be expressed as
\be\label{4.3}
\eta_N = \left. \frac{g \, B_1(\lambda)}{1 - g \, B_2(\lambda)} \, \right|_{g = k^{d-2} G, \lambda = \Lambda/k^2}
\ee
with the abbreviations
\be\label{4.4}
\begin{split}
B_1(\lambda) & \equiv \frac{1}{3} (4 \pi)^{1-d/2} \, \bigg[ d(d+1) \Phi^1_{d/2-1}(-2\lambda) - 6d(d-1) \Phi^2_{d/2}(-2\lambda) \\
& \qquad \qquad \qquad - 4d \Phi^1_{d/2-1}(0) - 24 \Phi^2_{d-2}(0) \bigg] \, , \\
B_2(\lambda) & \equiv -\frac{1}{6} (4 \pi)^{1-d/2} \left[ d(d+1) \tilde{\Phi}^1_{d/2-1}(-2\lambda) -6d(d-1) \tilde{\Phi}^2_{d/2}(-2\lambda) \right] \, . 
\end{split}
\ee
The ``threshold functions'' $\Phi^p_n$ and $\tilde{\Phi}^p_n$ are given by certain integrals which depend on the cutoff scheme, i.e., on $R_k(p^2)$. In Appendix \ref{apA} we discuss them for the cutoff schemes used in the present paper: the sharp cutoff \cite{frank1}, the exponential cutoff \cite{mr} and Litim's optimized cutoff \cite{optcutoff}. Note that the system of differential equations \eqref{1.auto} is an autonomous one, while \eqref{4.1} is not.

In this paper we shall solve \eqref{4.1} for $d=4$ numerically and use the resulting RG trajectories $(G(k), \Lambda(k))$ for the improvement procedure outlined in Section \ref{sect2}. Since the cosmological equations contain matter we should, in principle, use the beta-functions of gravity coupled to the corresponding matter system. Since we are interested in a qualitative understanding only, we shall use the flow equations of pure gravity, however. As in ref.\ \cite{h3}, our analysis is based upon the explicit assumption that the matter fields do not change the gross qualitative features of the pure gravity RG flow. Unless we know with certainty what the matter fields in Nature are we cannot anyhow decide on a theoretical basis whether or not this assumption is really correct.

\begin{section}{Cosmology on theory space}
The cosmological evolution with respect to the Robertson-Walker time $t$ is related to a scale-evolution via the cutoff identification $k = k(t)$. Therefore we may think of the history of the universe as a curve in the truncated theory space, $k \mapsto (g(k), \lambda(k))$.

Remarkably, certain properties of the universe at a given $t$ depend only on the point $(g,\lambda)$ in theory space where the universe happens to ``sit'' at that time, but not on $k$  or on the form of the trajectory. Examples include the matter and vacuum energy densities (divided by the critical density) and the deceleration parameter.

The present section is devoted to those properties of the RG improved cosmology which are directly related to the theory space (the $g$-$\lambda$-plane) and can be analyzed without first solving for the RG flow.

There exist two curves on the $g$-$\lambda$-plane which are important for a qualitative understanding of the improved cosmologies: the ``$\Omega$-line'' at which $\beta_\Lambda = 0$, and a line on which $\eta_N$ diverges.
\begin{subsection}{The $\Omega$-line}
By combining eqs.\ \eqref{2.4b} and \eqref{4.1} we can express the matter energy density directly in terms of beta-functions:
\be\label{10.1}
\rho(t(k)) = - \, \frac{1}{8 \pi} \, \frac{\beta_\Lambda(G(k), \Lambda(k), k)}{\beta_G(G(k), \Lambda(k), k)} \, .
\ee
Let us apply this formula to the class of cosmologies defined by the following properties: (i) They are not re-contracting, i.e., at least for $t \rightarrow \infty$ their scale factor $a(t)$ increases monotonically with $t$, and $\rho \propto 1/a^{3+3w}$ decreases correspondingly. (ii) The beta-function $\beta_G$ does not diverge during the evolution. We shall argue later on that the cosmology realized in Nature is likely to belong to this class.

For these cosmologies we have $a(t \rightarrow \infty) \rightarrow \infty$ and $\rho(t \rightarrow \infty) \rightarrow 0$. Hence their formal endpoints have $\rho = 0$ at ``$t = \infty$'' due to an eternal dilution of matter. In view of eq.\ \eqref{10.1} this means that $\beta_\Lambda$ vanishes for $t \rightarrow \infty$ since, by assumption, $\beta_G$ is always finite:
\be\label{10.2}
\lim_{t \rightarrow \infty} \beta_\Lambda \Big( G(k(t)), \Lambda(k(t)), k(t) \Big) = 0 \, .
\ee
In order to analyze the implication of this condition we first consider the more general equation
\be\label{10.3}
\beta_{\Lambda} \Big( G(k), \Lambda(k), k \Big) = 0
\ee
where no cutoff identification is invoked and $k$ is considered the independent variable. Remarkably, when re-expressed in terms of the dimensionless couplings $g \equiv k^2 G$ and $\lambda \equiv \Lambda / k^2$, eq.\ \eqref{10.3} becomes $k$-independent. (Here and in the following we specialize to $d=4$.) Using eq.\ \eqref{4.2b}, we see that $\beta_{\Lambda} = 0$ is equivalent to
\be\label{4.10}
\eta_N(g, \lambda) \, \lambda + \frac{1}{2 \pi} \, g \, \bigg[10 \, \Phi^1_{2}(-2 \lambda) 
 - 8 \Phi^1_{2}(0) - 5 \, \eta_N \tilde{\Phi}^1_{2}(-2 \, \lambda) \bigg] = 0 \, .  
\ee
 This provides a condition on $g$ and $\lambda$. It defines a curve on the $g$-$\lambda$-plane which we shall refer to as the ``$\Omega$-line''. It is the locus of all possible endpoints (``$\Omega$-points'') for the RG trajectories $k \mapsto (g(k), \lambda(k))$ belonging to the cosmologies with eternal expansion and $\beta_G < \infty$.

What will turn out crucial for the cosmologies of this class is that the $\Omega$-line coincides by no means with the boundary of the theory space, but is a well defined, albeit scheme dependent curve which lies at least partly in its interior. In fig.\ \ref{eins} it is plotted for the three cutoffs discussed in Appendix A.
\begin{figure}[t]
\renewcommand{\baselinestretch}{1}
\begin{center}
\leavevmode
\epsfxsize=0.7\textwidth
\epsffile{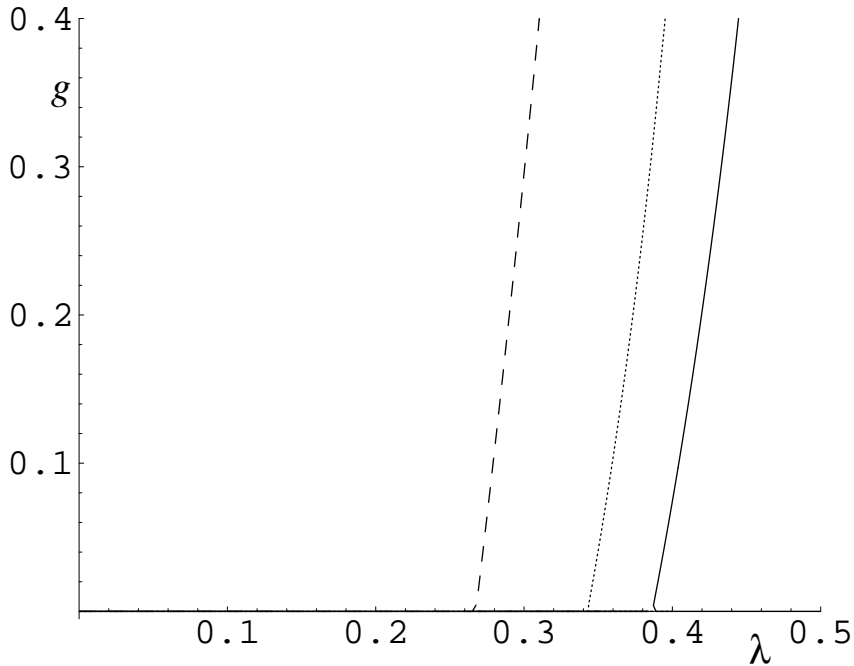} \,  
\end{center}
\parbox[c]{\textwidth}{\caption{\label{eins}{\footnotesize The $\Omega$-line for the optimized cutoff (dashed line), the exponential cutoff with $s=2$ (dotted line) and the sharp cutoff (solid line). The $\Omega$-line prevents cosmological solutions to run into the line $\lambda = 1/2$.}}}
\end{figure}

Let us assume we apply the method outlined in Section \ref{sect2} to a RG trajectory of Type IIIa such as the one depicted in fig.\ \ref{null}. Then it can happen that even after an {\it infinite} cosmological time the RG trajectory on the $g$-$\lambda$-plane does not reach the regime of possibly strong IR effects ($\lambda \lesssim 1/2$) but rather stops at some point on the $\Omega$-line. In this case the $\Omega$-line is ``screening'' the region of strong IR renormalizations. We shall see that, because of this mechanism, there are cosmologies which would {\it never} experience the conjectured strong IR quantum effects and remain essentially classical for $t \rightarrow \infty$.

However, the situation is somewhat involved since for most cutoff schemes the actual boundary of theory space is not precisely the straight line at $\lambda = 1/2$, but rather a complicated curve slightly left of it on which $\eta_N$ diverges already. (The sharp cutoff is an exception in the sense that $\eta_N$ diverges at $\lambda = 1/2$ only.) Therefore, in order to find out whether there can be a cosmological era with non-trivial IR effects, we must know the relative position of the $\Omega$-line and the boundary line with $|\eta_N| = \infty$. This will be the topic of the next subsection.
\end{subsection}
\begin{subsection}{Diverging anomalous dimension}
Eqs.\ (\ref{4.2a},b) show that $G$ and $\Lambda$ are strongly renormalized when the anomalous dimension $\eta_N$ is large. This happens close to certain curves on theory space on which $\eta_N$ diverges. As long as $|\eta_N|$ is only moderately large we tend to believe the predictions of the Einstein-Hilbert truncation. 

The anomalous dimension \eqref{4.3} can diverge in two ways:
\begin{enumerate}
\item The numerator of the RHS of \eqref{4.3} can diverge, $g B_1(\lambda) \rightarrow \infty$. This happens only along the line $\lambda = 1/2$.
\item The denominator on the RHS of \eqref{4.3} can vanish, $1-g B_2(\lambda) = 0$. 
\end{enumerate}
Analyzing the second possibility for the cutoff schemes from Appendix \ref{apA} we find that, for the sharp cutoff, this condition is never fulfilled for $g \ge 0$. For the exponential and the optimized cutoff, however, it is satisfied along a curve on the $g$-$\lambda$-plane which lies to the left of $\lambda = 1/2$ (for $g \ge 0$). This line can be found numerically; it is displayed in fig.\ \ref{einsa}.
\begin{figure}[t]
\renewcommand{\baselinestretch}{1}
\epsfxsize=0.48\textwidth
\begin{center}
\leavevmode
\epsffile{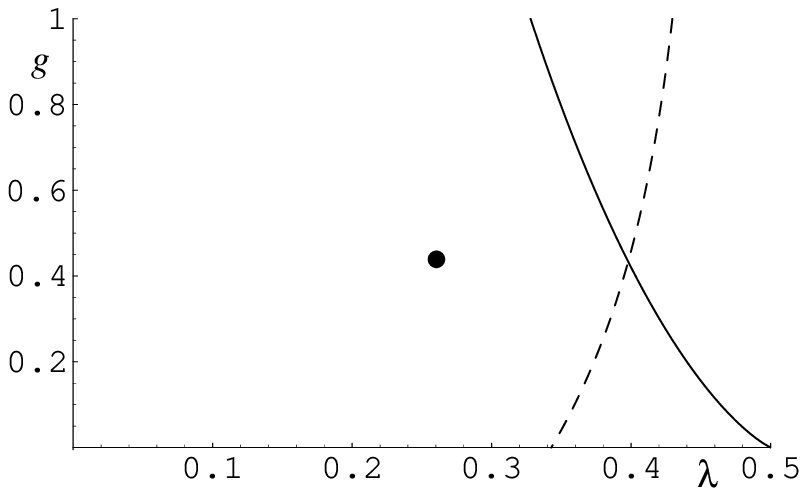} \,  
\epsfxsize=0.48\textwidth
\epsffile{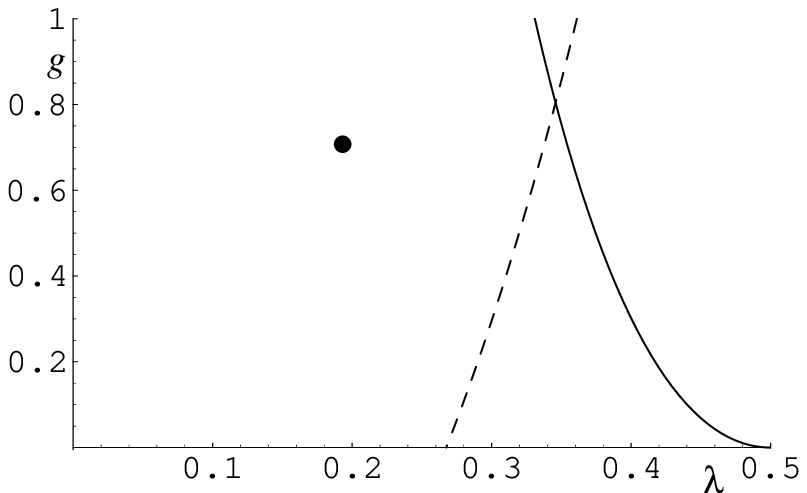}
\end{center}
\parbox[c]{\textwidth}{\caption{\label{einsa}{\footnotesize The lines of diverging $\eta_N$ for the exponential cutoff with $s=2$ (left diagram) and the optimized cutoff (right diagram). The solid lines correspond to the locus where $\eta_N$ diverges, while the dashed lines are the corresponding $\Omega$-lines. The location of the NGFP is indicated by the dot.}}}
\end{figure}

As a result the boundary of the theory space is the straight line $\lambda = 1/2$ for the sharp cutoff, and the curved lines of fig.\ \ref{einsa} for the other two cutoffs. On these boundaries, $\eta_N \rightarrow - \infty$, and no RG trajectory can be integrated beyond them. (See also \cite{litimneu}.)

In fig.\ \ref{einsa} the solid lines correspond to the locus where $\eta_N$ diverges, while the dashed lines indicate the position of the $\Omega$-line. For the Type IIIa trajectories coming from the left, the $\Omega$-line generically screens the singularity at $\lambda = 1/2$, but it does not provide a general screening mechanism for the other divergences which arise from a vanishing denominator in $\eta_N$. We observe that for a certain value $g = g_{\rm int}$ the $\Omega$-line and the locus of $\eta_N$-singularities  intersect. If $g > g_{\rm int}$ the $\eta_N$-singularities occur at smaller values of $\lambda$ than the $\Omega$-line. In this regime we find no screening of the $\eta_N$-singularity.

 For $g < g_{\rm int}$, however, the situation reverses and the $\Omega$-line occurs at smaller values of $\lambda$ than the $\eta_N$ singularity. In this case, both types of divergences in $\eta_N$, the ones coming from $\lambda = 1/2$ and the vanishing of the denominator in $\eta_N$ are shielded for all three cutoff schemes. This implies that a Type IIIa trajectory which is  sufficiently close to the $g = 0$-axis in the IR will {\it always} be stopped at the $\Omega$-line before the anomalous dimension $\eta_N$ diverges, independently of the cutoff scheme employed. In view of the tiny value $g = \cO(10^{-120})$ for $\lambda = \cO(1/2)$ this should in particular be the case for the trajectory realized in Nature.
\end{subsection}
\begin{subsection}{Energy densities and deceleration parameter}
In order to study the evolution of the energy densities associated to the matter and the cosmological constant, it is useful to introduce the relative matter and vacuum energy densities: 
\be\label{9.1}
\rho_{\rm crit} \equiv \frac{3 H^2}{8 \pi G} \, , \quad \Omega_M \equiv \frac{\rho}{\rho_{\rm crit}} = - \, \frac{1}{3 H^2} \, \frac{\Lambda^{\prime}}{G^{\prime}/G} \, , \quad \Omega_{\Lambda} \equiv \frac{\rho_{\Lambda}}{\rho_{\rm crit}} =  \frac{\Lambda}{3 H^2}  \, .
\ee
We assume a spatially flat universe ($K = 0$) and therefore have $\Omega_{\rm tot} \equiv \Omega_M + \Omega_{\Lambda} = 1$ as a consequence of the system (\ref{2.1a},b,c) \cite{cosmo1}. Using the Friedmann equation \eqref{2.1a}, we can eliminate the Hubble parameter from the equations for $\Omega_M$ and $\Omega_\Lambda$:
\be\label{9.2}
\frac{1}{\Omega_M} = 1 - \, \frac{G^{\prime} / G}{\Lambda^{\prime} / \Lambda} \; , \quad 
\frac{1}{\Omega_\Lambda} = 1 - \, \frac{\Lambda^{\prime} / \Lambda}{G^{\prime} / G} \, .
\ee
This motivates defining the function
\be\label{9.3}
Y(k) \equiv \frac{\Lambda^{\prime} / \Lambda}{G^{\prime} / G} \, ,
\ee
which completely determines the evolution of $\Omega_M$ and $\Omega_\Lambda$:
\be\label{9.4}
\Omega_M = - \, \frac{Y(k)}{1 - Y(k)} \, , \quad \Omega_\Lambda = \frac{1}{1-Y(k)} \, .
\ee
Here $Y(k)$ is understood as being evaluated along a particular solution of the RG equations. The above relations imply that, for $Y(k) = 0$, the total energy density is completely dominated by the cosmological constant ($\Omega_M = 0, \Omega_\Lambda = 1$), while $Y(k) = \pm \infty$ corresponds to complete matter domination ($\Omega_M = 1, \Omega_\Lambda = 0$).

These relations can be used to express the deceleration parameter
\be\label{9.5a}
q(t) \equiv - \frac{\add \, a}{\ad^2}
\ee
in terms of $Y(k)$. From the improved system (\ref{2.1a},b,c) one derives \cite{cosmo2} that, as in the classical case,
\be\label{9.5b}
q = \tfrac{1}{2} \, (3 w + 1) \, \Omega_M - \Omega_{\Lambda} \, .
\ee
Substituting \eqref{9.4} we obtain
\be\label{9.5c}
q = \frac{1}{Y-1} \left[ \frac{3w+1}{2} Y + 1 \right] \, .
\ee

Writing
\be\label{9.5}
Y(k) = y \Big( g(k), \lambda(k) \Big)
\ee
in terms of the dimensionless couplings $g, \lambda$ it turns out that
\be\label{9.6}
y(g,\lambda) = 1 + \frac{1}{2 \pi \eta_N(g,\lambda)} \, \frac{g}{\lambda} \, \left[ 10 \, \Phi^1_{2}(-2 \lambda) - 8 \Phi^1_{2}(0) - 5 \, \eta_N \tilde{\Phi}^1_{2}(-2 \, \lambda) \right]
\ee
has no explicit $k$-dependence. By definition, $Y$ is a function of $k$, while $y$ is a function of $g$ and $\lambda$. Hence $\Omega_M$, $\Omega_\Lambda$, and $q$ are completely determined by the values of 
$g$ and $\lambda$:
\be\label{4.y}
\Omega_M = - \, \frac{y(g,\lambda)}{1-y(g,\lambda)} \, , \; \; \; \Omega_\Lambda = \frac{1}{1-y(g,\lambda)} \, , \; \; \; q = \frac{1+(3w+1)\, y(g,\lambda)/2}{y(g,\lambda)-1} \, .
\ee
Thus it is important to know the properties of the function $y(g,\lambda)$.

We start by comparing eq.\ \eqref{9.6} to the condition determining the $\Omega$-line, eq.\ \eqref{4.10}. This shows that  $y(g,\lambda)=0$ at the $\Omega$-line, so that the energy density of a universe on this line is completely dominated by the cosmological constant, $\Omega_\Lambda = 1, \Omega_M = 0$.

In our further analysis we first focus on the $g=0$-limit of $y(g,\lambda)$. This is motivated by the fact that the trajectory realized in Nature has $g \ll 1$ so that this limit provides a good approximation. Setting $g=0$ in \eqref{9.6} we then obtain:
\be\label{9.7}
y_{0}(\lambda) \equiv \lim_{g \rightarrow 0} \, y(g, \lambda) = 1 + \frac{1}{2 \pi \lambda B_1(\lambda)} \left[ 10 \, \Phi^1_{2}(-2 \lambda) - 8 \Phi^1_{2}(0) \right] \, .
\ee
Using the sharp cutoff, this function is displayed in fig.\ \ref{nine}.
\begin{figure}[t]
\renewcommand{\baselinestretch}{1}
\epsfxsize=0.48\textwidth
\begin{center}
\leavevmode
\epsffile{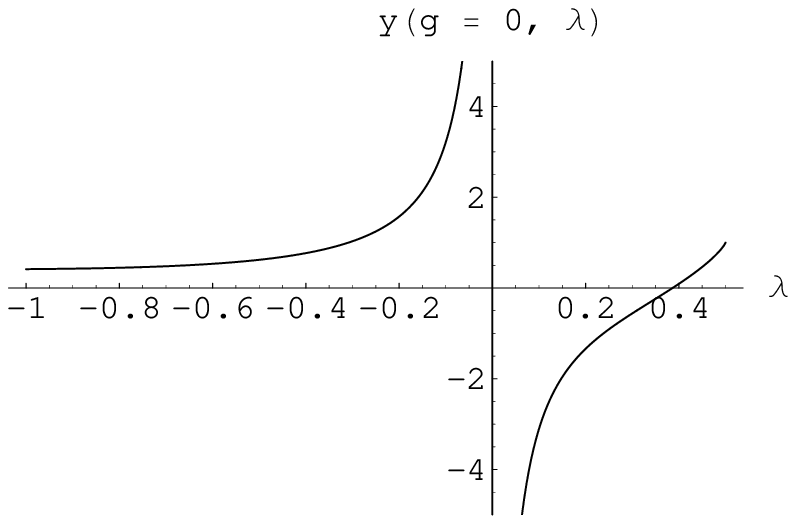} \,  
\epsfxsize=0.48\textwidth
\epsffile{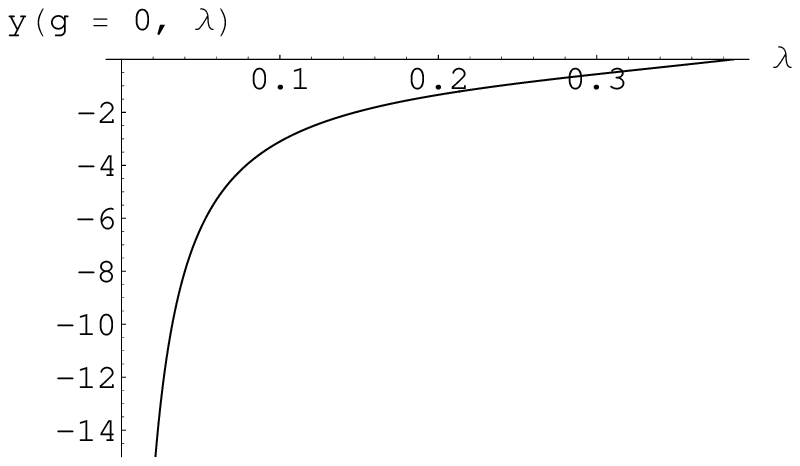}
\end{center}
\parbox[c]{\textwidth}{\caption{\label{nine}{\footnotesize The function $y_0(\lambda)$. Its zero corresponds to the $\Omega$-line at $g=0$. The right figure illustrates that near this zero $y_0(\lambda)$ is an approximately linear function.}}}
\end{figure}
Looking at the left diagram, we observe that the domain of definition of $y_0{}$, ($-\infty < \lambda \le 1/2$), contains the following special points and intervals: 
\be 
\begin{split}
\lambda \rightarrow - \infty \quad & \Rightarrow \quad y_0(\lambda) \searrow  1 \, , \\ 
- \infty < \lambda < 0       \quad & \Rightarrow \quad y_0(\lambda) > 1 \, , \\ 
\lambda = 0^+               \quad & \Rightarrow  \quad y_0(\lambda) = - \infty \, , \\ 
0 < \lambda \le \lambda_{\Omega-\mbox{line}} \quad &\Rightarrow \quad y_0(\lambda) \le 0 \, ,  \\  
\lambda_{\Omega-\mbox{line}} < \lambda < 1/2 \quad &\Rightarrow \quad 0 < y_0(\lambda) < 1 \, , \\ 
\lambda = 1/2                               \quad &\Rightarrow \quad y_0(\lambda) = 1 \, .
\end{split}
\ee
Fig.\ \ref{ten} shows the full function $y(g, \lambda)$ on the $g$-$\lambda$-plane. This figure illustrates that the behavior described above also extends to the region where $g>0$ so that the discussion given below will also apply to a general trajectory with $g>0$. In particular, close to the $\Omega$-line, $y$ is an approximately linear function of $\lambda$.

\begin{figure}[t]
\renewcommand{\baselinestretch}{1}
\epsfxsize=0.45\textwidth
\begin{center}
\leavevmode
\epsffile{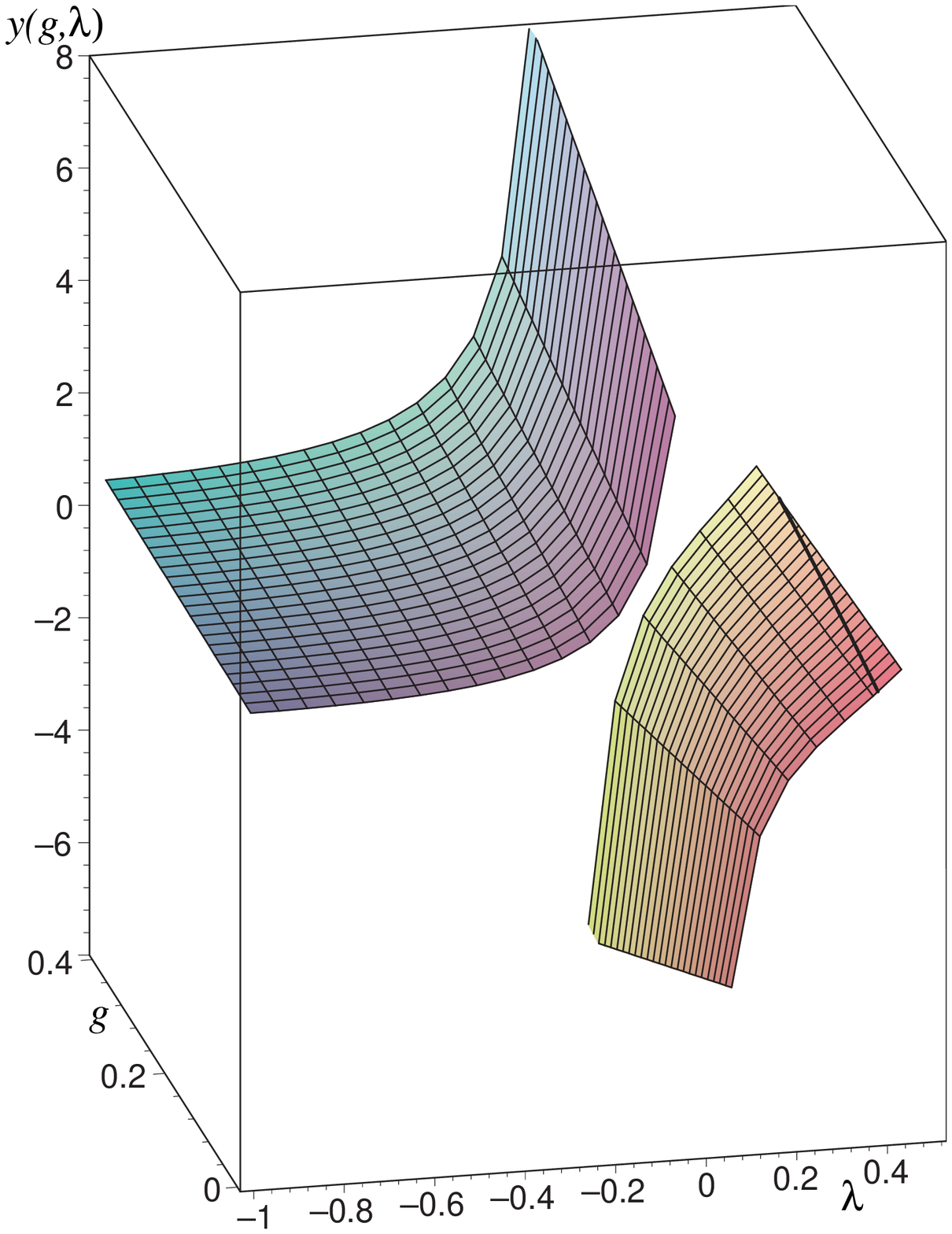} \,  
\epsfxsize=0.45\textwidth
\epsffile{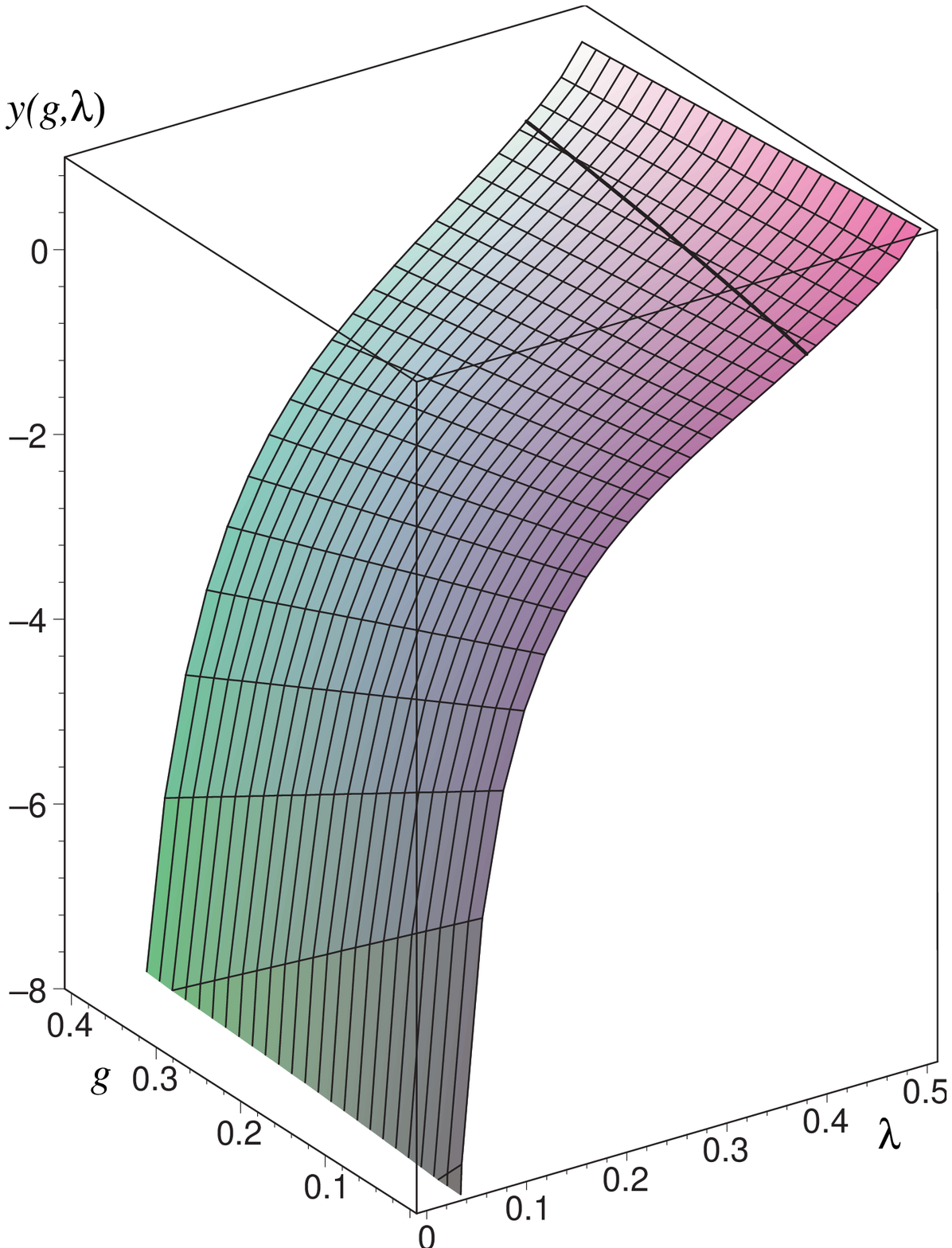}
\end{center}
\parbox[c]{\textwidth}{\caption{\label{ten}{\footnotesize The function $y(g, \lambda)$. The solid black line represents the $\Omega$-line where $y(g, \lambda) = 0$. Note that $y(g, \lambda \searrow  0) \rightarrow - \infty$ which corresponds to a completely matter dominated universe.}}}
\end{figure}

In terms of $\Omega_M$ and $\Omega_\Lambda$, the features of $y$ found above have the following interpretation. The region $- \infty < \lambda < 0$ is associated to the IR behavior of RG-trajectories of Type Ia. A cosmology evolving along such a trajectory will asymptote to $\Omega_\Lambda \rightarrow - \infty$, $\Omega_M \rightarrow + \infty$ with the sum  $\Omega_\Lambda + \Omega_M = 1$. 

Next we have the separatrix which, in the IR, ends at the GFP $g_* = 0, \lambda_* = 0$. At this point $y_0(\lambda = 0^+) = - \infty$, so that the cosmology asymptotes to $\Omega_M = 1, \Omega_\Lambda = 0$. The resulting universe will then be completely matter dominated at late times. 

In the region $0 < \lambda \le \lambda_{\Omega-\mbox{line}}$, we find the RG-trajectories of Type IIIa, discussed in the previous section. In this case the cosmological evolution ends at the $\Omega$-line where $y(g, \lambda) = 0$. Hence these cosmologies lead to a domination of the cosmological constant at late times: $\Omega_M = 0, \Omega_\Lambda = 1$. 

An interesting property of this latter region is that it splits into two rather different parts. In the first part, where $y(g, \lambda)$ increases rapidly, we observe matter domination while in the linear regime $\Omega_\Lambda$ dominates. 

In the region $\lambda_{\Omega-\mbox{line}} < \lambda < 1/2$ we have $\Omega_\Lambda > 1$ and $\Omega_M < 0$, which is (probably) unphysical.

For later use we note the value of the function $y$ at the NGFP $(g_*, \lambda_*)$:
\be\label{4.yfp}
y(g_*, \lambda_*) = -1
\ee
This result is most easily proven by simply inserting $\Lambda(k) = \lambda_* k^2$ and $G(k) = g_* / k^2$ into eq.\ \eqref{9.3}. By virtue of eq.\ \eqref{4.y}, this implies that $\Omega_M^* = \Omega_\Lambda^* = 1/2$ and $q^* = (3w-1)/4$ at the fixed point.
\end{subsection}
\end{section}
\begin{section}{Evolution in time and scale}
The proper way of thinking about an RG improved cosmological history is to visualize it as a curve on theory space, the RG trajectory, which, by means of eqs.\ (\ref{2.1a},b,c), induces both a Robertson-Walker scale factor $a(t)$ and a cutoff identification $k = k(t)$. In this section we ``switch on'' this $k$- or $t$-evolution and study, mostly by numerical methods, the coupled system (\ref{2.1a},b,c) after having first obtained a trajectory $(g(k), \lambda(k))$, or rather $(G(k), \Lambda(k))$, by solving the flow equations \eqref{4.1}. Because of their special relevance we shall consider trajectories of Type IIIa only. Since the crucial qualitative features such as the existence of an $\Omega$-line are the same in all cutoff schemes, we shall use the technically convenient sharp cutoff throughout. 

\begin{subsection}{Initial conditions and RG trajectories}
To investigate the properties of the solutions $(g(k),\lambda(k))$ numerically we choose our initial conditions for the dimensionless couplings $g, \lambda$ along the line connecting the turning points $(g_T,\lambda_T)$ of the trajectories. (See fig.\ \ref{zwei}.) For the sharp cutoff and close to the GFP this line is given by \cite{h3}
\be\label{6.1}
\lambda_T = \frac{\varphi_2}{2 \pi} \, g_T \, .
\ee
For the numerics we supplement this relation with
\be\label{6.2}
k_{\rm init} = 1 \, , \quad t_{\rm init} = 0 \, , \quad a_{\rm init} = 1 \, . 
\ee
This gives rise to a one-parameter family of cosmological solutions which, loosely speaking, are characterized by their distance to the GFP at $\lambda = 0, g = 0$. By definition, the trajectories pass the turning point at the scale $k_T$, i.e., $g(k_T) = g_T$ and $\lambda(k_T) = \lambda_T$. Thus, adopting the relations \eqref{6.2} amounts to expressing all dimensionful quantities in terms of appropriate powers of $k_T$. (In this parameterization the Big Bang occurs at $t<0$.)

Starting with these initial values we evolve the solutions in two directions:
\begin{itemize}
\item Towards the UV ($k > k_T$) where the RG-Flow will be attracted towards the NGFP. 
\item Towards the IR ($k < k_T$) where the cosmology will run towards the $\Omega$-line. 
\end{itemize}
For explicitness we consider the three sample solutions with
\be\label{6.3}
g_T = 10^{-1} \; , \quad g_T = 10^{-2} \; , \; \mbox{and} \quad g_T = 10^{-3} \,.
\ee
These examples are sufficient to illustrate the general trend for decreasing $g_T$ and to understand the qualitative properties of the RG-trajectory realized in Nature which has an extremely tiny $g_T$.
The RG-trajectories obtained by numerically solving \eqref{4.1} with these initial conditions are shown in fig.\ \ref{zwei}.
\begin{figure}[t]
\renewcommand{\baselinestretch}{1}
\begin{center}
\leavevmode
\epsfxsize=0.6\textwidth
\epsffile{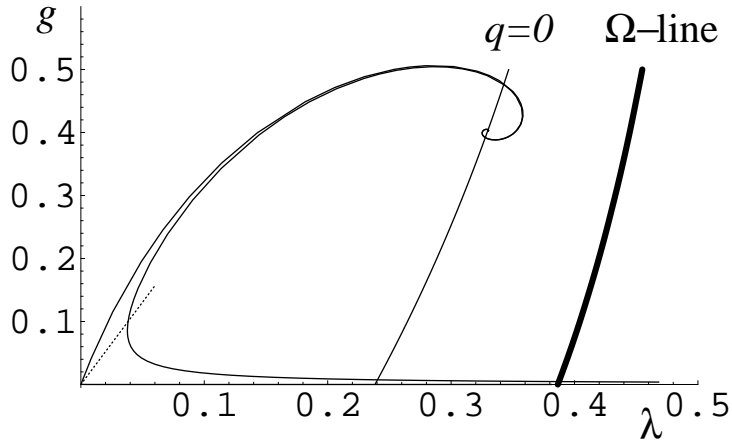} \,  
\end{center}
\parbox[c]{\textwidth}{\caption{\label{zwei}{\footnotesize RG-trajectories with the initial conditions given in \eqref{6.3}. The dotted line is the locus of the turning points which  separate their UV and IR parts. The two almost vertical lines are the $q=0$-locus for $w=1/3$ and the $\Omega$-line.}}}
\end{figure}
We see that decreasing $g_T$ results in squeezing the solution into the corner between the separatrix (connecting the NGFP and the GFP) and the $g = 0$-axis. In fact, within the resolution of fig.\ \ref{zwei} the two trajectories with $g_T = 10^{-2}$ and $g_T = 10^{-3}$ run virtually on top of the separatrix (for $k > k_T$) and the horizontal axis (for $k<k_T$).

All three trajectories are of Type IIIa and terminate at the boundary $\lambda = 1/2$. For $g_T = 10^{-1}, 10^{-2}$, and $10^{-3}$ the termination scales are found to be $k_{\rm term}/k_T = 0.178, 0.0612$, and $0.0196$, respectively. These numbers are consistent with the rough estimate $k_{\rm term}/k_T = \cO(\sqrt{g_T})$ derived in \cite{h3}.\footnote{The proportionality constant is $k_{\rm term}/k_T \approx 0.612 \sqrt{g_T}$. The first case $g_T = 10^{-1}$ does not quite fit into the pattern, since this trajectory is not yet sufficiently deep in the GR regime.} They confirm that the closer the trajectory approaches the GFP the later (in $k$) it terminates.  
\end{subsection}
\begin{subsection}{UV vs.\ IR branch}
The two branches of the trajectory, $k > k_T$ and $k < k_T$, give rise to two corresponding branches of the cosmological solution which we shall refer to as the ``UV cosmology'' and ``IR cosmology'', respectively. In the UV cosmology we are going to use the equation of state with $w = 1/3$ (``radiation''), while we employ $w = 0$ (``dust'') for the IR cosmology. In reality the universe has passed the scale $k = k_T$ well inside the radiation dominated era, but since we are mostly interested in qualitative features of RG improved cosmologies this aspect is inessential.

Next we shall apply the method described in Section \ref{sect2} to the trajectories of the previous subsection and discuss their UV- and IR-branches in turn. In all examples the Hubble parameter is calculated using eq.\ \eqref{B.5} derived in Appendix \ref{AppB}, and the deceleration parameter is determined with the help of eq.\ \eqref{4.y}.


%
\begin{figure}[t]
\renewcommand{\baselinestretch}{1}
\begin{center}
\leavevmode
\epsfxsize=0.6\textwidth
\epsffile{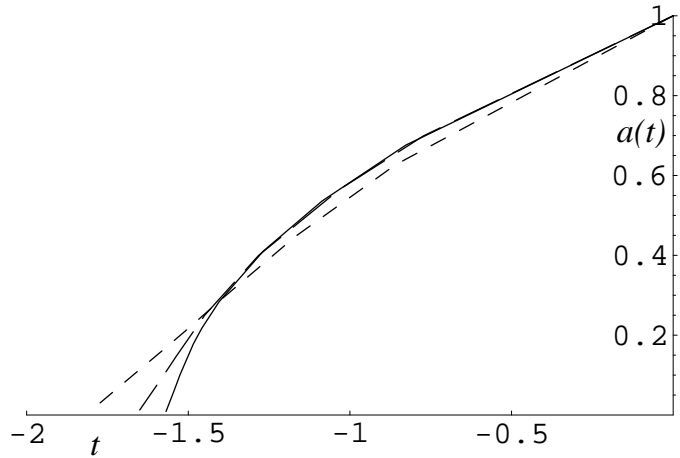} \,  
\end{center}
\parbox[c]{\textwidth}{\caption{\label{auv}{\footnotesize The scale factor $a(t)$ for the UV cosmologies arising from $g_T = 10^{-1}$ (short dashes), $g_T = 10^{-2}$ (long dashes) and $g_T = 10^{-3}$ (solid line).}}}
\end{figure}
\begin{figure}[t]
\renewcommand{\baselinestretch}{1}
\epsfxsize=0.48\textwidth
\begin{center}
\leavevmode
\epsffile{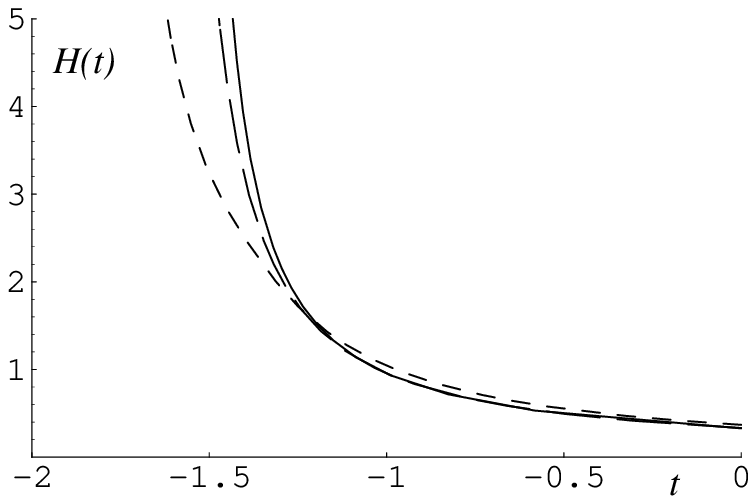}   
\epsfxsize=0.48\textwidth
\epsffile{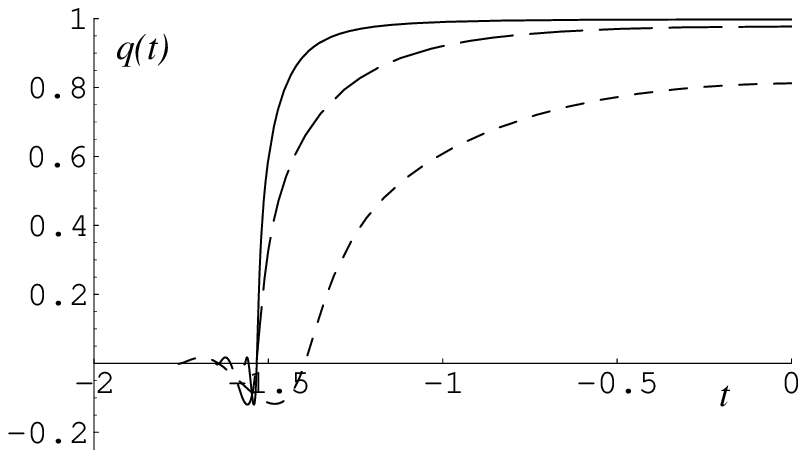}
\end{center}
\parbox[c]{\textwidth}{\caption{\label{sieben}{\footnotesize $H(t)$ and $q(t)$ for the UV cosmologies starting with  $g_T = 10^{-1}$ (short dashes), $g_T = 10^{-2}$ (long dashes) and $g_T = 10^{-3}$ (solid line).}}}
\end{figure}
\end{subsection}
\begin{subsection}{RG-improved UV cosmologies}
The numerical results for the UV cosmologies resulting from the $g_T$-values \eqref{6.3} are displayed in figs.\ \ref{auv} - \ref{dreizehn}. Fig.\ \ref{auv} shows the scale factor $a(t)$, and fig.\ \ref{sieben} the corresponding Hubble constant $H \equiv \ad/a$ and deceleration parameter $q$. In fig.\ \ref{sechs} Newton's constant $G(t) \equiv G(k(t))$ and likewise the cosmological constant is plotted as a function of time. Here $t$ is measured in units of $k_T^{-1}$, and the origin $t=0$ corresponds to $k = k_T$.
\begin{subsubsection}{The regimes of the UV cosmologies}
\label{5.3.1}
We observe that the cosmologies possess an initial singularity (``Big Bang'') at a time $t_B < 0$ where $a(t_B) = 0$ and $H(t \searrow t_B) \rightarrow \infty$. For $t > t_B$ we can distinguish the following regimes:\\
{\bf (A) The NGFP regime \\}
Immediately after the Big Bang, for $k \gg m_{\rm Pl}$, the RG trajectory is well approximated by the constant functions $g(k) = g_*$ and $\lambda(k) = \lambda_*$. They correspond to the $k$-dependence \eqref{1.1} for which the system (\ref{2.1a},b,c) can be solved analytically. This yields the fixed point solution found in ref.\ \cite{cosmo1}. For $w = 1/3$ it yields $ a \propto (t-t_B)$, $H = 1/(t-t_B)$, $q = 0$, $G \propto (t-t_B)^2$, and $\Lambda \propto 1/(t-t_B)^2$. The numerical solutions do indeed show this behavior close to $t_B$. In particular, $G$ vanishes and $\Lambda$ diverges at the Big Bang.
\begin{figure}[t]
\renewcommand{\baselinestretch}{1}
\epsfxsize=0.49\textwidth
\begin{center}
\leavevmode
\epsffile{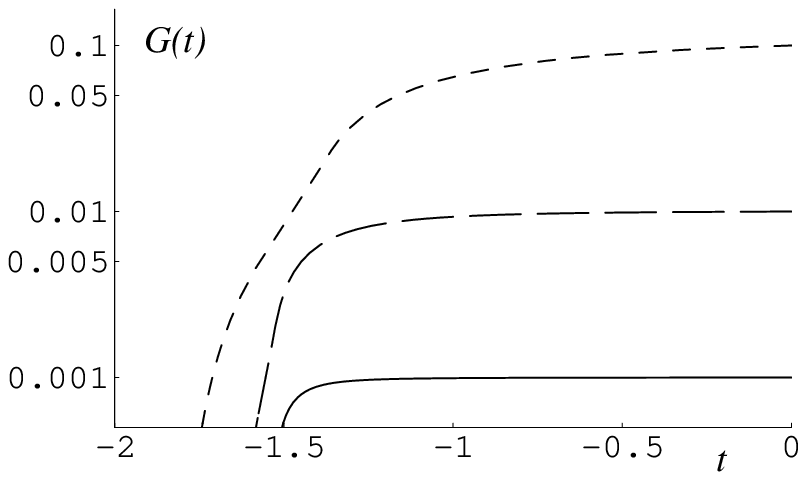}   
\epsfxsize=0.49\textwidth
\epsffile{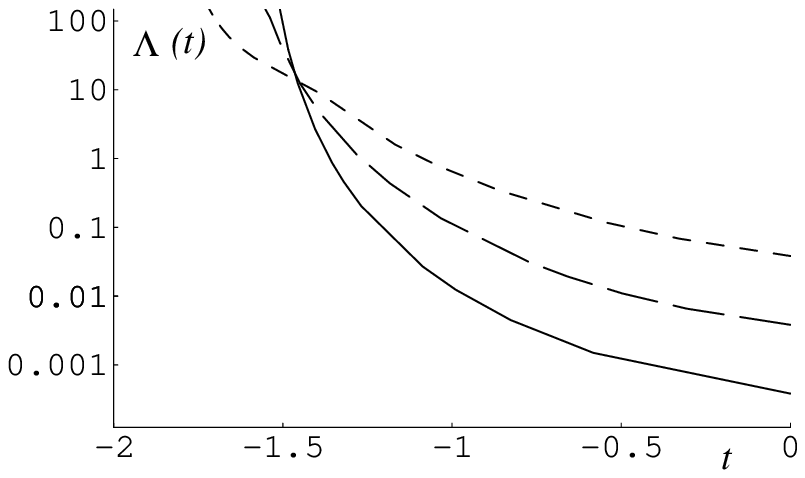}
\end{center}
\parbox[c]{\textwidth}{\caption{\label{sechs}{\footnotesize The functions $G(t), \Lambda(t)$ for the UV cosmologies arising from the initial conditions $g_T = 10^{-1}$ (short dashes), $g_T = 10^{-2}$ (long dashes) and $g_T = 10^{-3}$ (solid line).}}}
\end{figure}

\noindent
{\bf (B) The linear regime of the NGFP\\}
At slightly later time, corresponding to smaller scales $k$, the RG flow can be linearized about the NGFP \cite{frank1,oliver1}. The linearized trajectories $(g(k), \lambda(k))$ are spirals about the NGFP, characterized by two critical exponents $\theta'$ and $\theta''$ \cite{frank1,oliver1}. The most prominent cosmological signature of this regime is the oscillatory behavior of $q(t)$ seen in fig.\ \ref{sieben} and, in a slightly magnified way, in fig.\ \ref{dreizehn}. Directly at the NGFP one has $q=0$; when the trajectory starts circling about the NGFP there are phases with both $q > 0$ (deceleration) and $q < 0$ (acceleration). In fact, there exists an infinite sequence of phases with either sign. When one approaches the Big Bang from above ($t \searrow t_B$), $q$ decreases from about $q \approx 1$, has a first zero, becomes negative, has another zero, becomes positive for a period shorter than the previous one, then again becomes negative for a time span shorter than the previous one, and so on. The Big Bang is approached by an infinite sequence of oscillations in $q$, of decreasing amplitude and decreasing duration.
\begin{figure}[t]
\renewcommand{\baselinestretch}{1}
\epsfxsize=0.65\textwidth
\begin{center}
\leavevmode
\epsffile{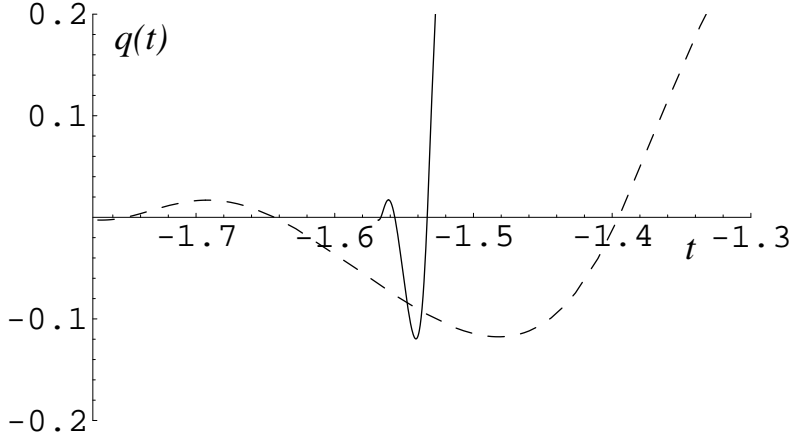}
\end{center}
\parbox[c]{\textwidth}{\caption{\label{dreizehn}{\footnotesize Oscillatory phases of the deceleration parameter $q(t)$ for the UV cosmologies starting with  $g_T = 10^{-1}$ (short dashes) and $g_T = 10^{-3}$ (solid line).}}}
\end{figure}

\noindent
{\bf (C) Crossover and linear regime of the GFP\\}
At still lower scales, or later times, the RG trajectories leave the linear regime of the NGFP and very quickly ``cross over'' towards the GFP. Linearizing about the GFP, the resulting cosmologies are again easily understood analytically. One finds \cite{h3} that approximately $G(k) = const$ and $\Lambda(k) = \Lambda(k_T) \, [1+(k/k_T)^4]/2$. If $k$ is related to $t$ in a monotonic way, we expect $G$ to stay constant and $\Lambda$ to decrease towards its value at $k_T$ (where it becomes constant, too). This is exactly what we observe in fig.\ \ref{sechs}. 

In the linear regime of the GFP, for the time $t < 0$, i.e., before the turning point of the trajectory, the cosmological constant is not important for the three solutions. At least for $g_T$ small they have $\Omega_\Lambda \approx 0$ and $\Omega_M \approx 1$ there, yielding $q \approx 1$. This is exactly the plateau-value of $q(t)$ which is approached in the second plot of fig.\ \ref{sieben}.
\end{subsubsection}
\begin{subsubsection}{Dynamical determination of $k = k(t)$}
As we explained in Section \ref{sect2}, our method determines the cutoff identification $k = k(t)$ dynamically in such a way that the field equations are integrable by construction, the risk being that the $k(t)$ thus obtained does not lend itself for a clear interpretation of the resulting coarse-grained cosmology.
\begin{figure}[t]
\renewcommand{\baselinestretch}{1}
\begin{center}
\leavevmode
\epsfxsize=0.49\textwidth
\epsffile{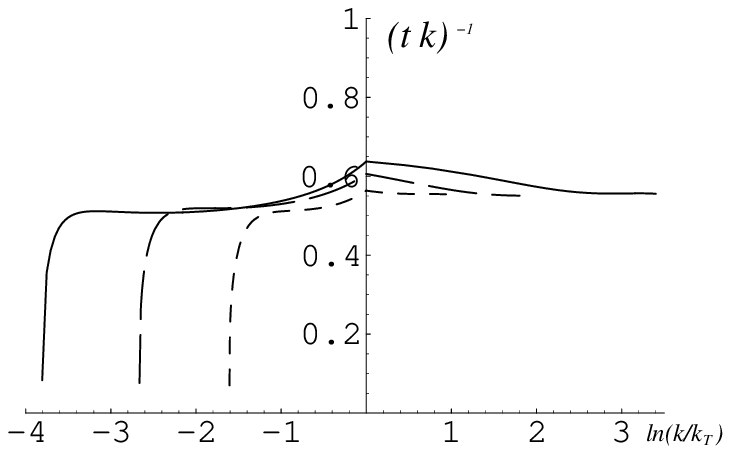}   
\epsfxsize=0.49\textwidth
\epsffile{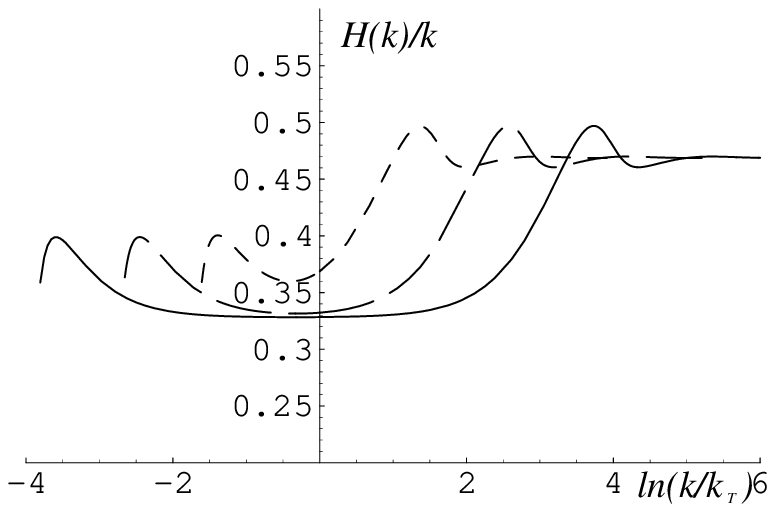}
\end{center}
\parbox[c]{\textwidth}{\caption{\label{acht}{\footnotesize $(t(k) k )^{-1}$ and $H(t(k))/k$ for the combined UV and IR cosmologies with  $g_T = 10^{-1}$ (short dashes), $g_T = 10^{-2}$ (long dashes) and $g_T = 10^{-3}$ (solid line).}}}
\end{figure}
In the first diagram of fig.\ \ref{acht} we display the relationship between $k$ and $t$ by plotting $(t(k) k)^{-1}$ as a function of $\ln(k/k_T)$. The motivation for this presentation is as follows. If this function is $k$-independent, $t(k) k = const$, we have $t(k) \propto 1/k$ or vice versa $k(t) \propto 1/t$ which is the rigid cutoff identification used in \cite{cosmo1,cosmo2}. In order to get the $(t(k) k )^{-1}$ vs. $\ln(k/k_T)$ plot, we have shifted $t(k)$ such that the Big Bang ($k = \infty$) corresponds to $t(k = \infty) = 0$.\footnote{This is different in figs.\ \ref{sieben} and \ref{sechs} where the UV cosmology corresponds to $t < 0$. At $t = 0 \Leftrightarrow  k = k_T = 1$, the expression $(t(k) k )^{-1}$ develops a pole, which is avoided by shifting the Big Bang to $t = 0$.} In an analogous fashion the second plot in fig.\ \ref{acht} shows $H(k)/k$ as a function of $k$; this function would be constant for the identification $k(t) \propto H(t)$. 

From fig.\ \ref{acht} we see that both $k \propto H(k)$ and $k \propto 1/t(k)$ provide a valid approximation to the true cutoff-identification in the UV domain ($k > k_T$) if one is interested in the overall qualitative features only. In this domain the approximation $k \propto 1/t(k)$ performs somewhat better than $k \propto H(k)$.
\end{subsubsection}
\begin{subsubsection}{Oscillatory inflation}
Finally we return to the phenomenon of the $q(t)$-oscillations in the NGFP-scaling regime which we described in \ref{5.3.1} (B). In order to analyze the sign-flips of the deceleration parameter we recall from \eqref{4.y} that $q = q(g, \lambda)$ can be regarded as a function on theory space, the $t$-dependence arising by inserting the trajectory with $k = k(t)$, i.e., $q(t) \equiv q(g(k(t)), \lambda(k(t)))$.

As for the function $q(g, \lambda)$, there are regions on the $g$-$\lambda$-plane where it is positive and others where it is negative, the common boundary being the ``$q = 0$-line'' shown in fig.\ \ref{zwei}. By virtue of \eqref{4.y} this line is given by the implicit condition $1+(3w+1)\, y(g,\lambda)/2 = 0$. To obtain the diagram in fig.\ \ref{zwei} the equation for $w = 1/3$, $y(g, \lambda) = -1$, was solved numerically. To the right (left) of the $q=0$-line one has $q < 0$ ($q > 0$).

Now it is important to observe that precisely for $w = 1/3$ the $q = 0$-line runs exactly through the NGFP. In fact, according to \eqref{4.yfp} the NGFP has $y(g_*, \lambda_*) = -1$, i.e., it lies on the $q=0$-line pertaining to $w = 1/3$. As a consequence, when the RG trajectory spirals around the NGFP it crosses the $q=0$-line an infinite number of times in either direction. This explains the $q$-oscillations about $q=0$ found in the numerical solutions.

Thus the RG-improvement predicts epochs of accelerated expansion (``inflation'') in the early universe and one might wonder whether there can be any relation to the traditional models of cosmological inflation. To analyze this question we compute the number of ``$e$-folds'' the universe expands during the periods with $q < 0$. We consider only the last such period because it has the longest duration and the most negative $q(t)$, see fig.\ \ref{dreizehn}. Note that even there $q(t)$ never becomes as small as $-1$ which would correspond to a de~Sitter-like phase.

We first derive a general formula which can be used to calculate the number of $e$-folds of expansion, $N$, occurring along any RG-trajectory. This formula will then be used to determine $N$ for the last period of accelerated expansion occurring before the solution leaves the linear regime of the NGFP (see fig.\ \ref{zwei}). 

Our starting point is the usual definition of $N$,
\be\label{7.1}
N \equiv \ln \left[ \frac{a(t_{\rm e})}{a(t_{\rm b})} \right] \, ,
\ee 
where $t_{\rm b}$ and $t_{\rm e}$ denote the cosmological times where the acceleration begins and ends, respectively. Using \eqref{2.4a} we can relate $N$ directly to the RG-trajectory:
\be\label{7.2}
N = \frac{1}{3+3w} \, \ln \left[ \frac{G^{\prime}_{\rm e}}{\Lambda^{\prime}_{\rm e}} \, \frac{\Lambda^{\prime}_{\rm b}}{G^{\prime}_{\rm b}} \right] \, .
\ee
Here and in the following the subscripts ${\rm b}$, ${\rm e}$ denote the corresponding quantity at the beginning and end of the acceleration period. Eq.\ \eqref{7.2} can be evaluated further by re-expressing the quantities $G^\prime$, $\Lambda^{\prime}$ using eq.\ \eqref{B.2}. Writing the result in terms of the dimensionless couplings $g, \lambda$ we find
\be\label{7.3}
N = \frac{4}{3+3w}  \, \ln\left[ \frac{k_{\rm b}}{k_{\rm e}} \right] + \frac{1}{3+3w} \ln \left[ \frac{A(g_{\rm e}, \lambda_{\rm e})}{A(g_{\rm b}, \lambda_{\rm b})} \right] \, ,
\ee 
where we define
\be\label{7.4}
A(g, \lambda) \equiv g \left[ \lambda + \frac{g}{2 \pi \eta^{\rm sc}} \left( -10 \, \ln(1-2\lambda) + 2 \, \zeta(3) - \tfrac{5}{2} \, \eta^{\rm sc} \right) \right]^{-1} \, .
\ee
In terms of the RG-time $\tau \equiv \ln(k/k_T)$, and for $w = 1/3$, this becomes simply
\be\label{7.5}
N = \tau_{\rm b} - \tau_{\rm e} + \frac{1}{4} \ln \left[ \frac{A(g_{\rm e}, \lambda_{\rm e} )}{A(g_{\rm b}, \lambda_{\rm b})} \right] \,. 
\ee 

Looking at fig.\ \ref{zwei} we see that the coordinates of the last and the second to last $q=0$-crossing, ($g_{\rm e}, \lambda_{\rm e}$) and ($g_{\rm b}, \lambda_{\rm b}$), are to a very good precision equal for all Type IIIa trajectories, so the last term in \eqref{7.5} gives rise to an almost ``universal'' contribution. 

We then evaluate $N$ along the last period of accelerated expansion, the semi-circle in fig.\ \ref{zwei}. The corresponding values are summarized in table \ref{three}.
\begin{table}[t]
\begin{tabular*}{\textwidth}{@{\extracolsep{\fill}} c||c|c||c|c||c } 
$g_T$ & $\tau_{\rm b}$ & $\tau_{\rm e}$ & $A_{\rm b}$ & $A_{\rm e}$ & $N$ \\ \hline 
$10^{-1}$ & $2.505$ & $1.642$ & $-1.196$ & $-1.390$ & $0.901$ \\
$10^{-2}$ & $3.710$ & $2.848$ & $-1.196$ & $-1.392$ & $0.900$ \\
$10^{-3}$ & $4.886$ & $4.004$ & $-1.195$ & $-1.392$ & $0.900$ \\
$10^{-10}$ & $12.811$ & $11.945$ & $-1.196$ & $-1.392$ & $0.900$ \\
\end{tabular*}
\renewcommand{\baselinestretch}{1}
\parbox[c]{\textwidth}{\caption{\label{three}{\footnotesize
Number of $e$-folds obtained in the last period of accelerated expansion when a Type IIIa trajectory is encircling the NGFP.}}}
\end{table}
From this table we observe that $A_{\rm b}$ and $A_{\rm e}$ are indeed independent of $g_T$ as suggested by fig.\ \ref{zwei}. Furthermore, we find that while the values of $\tau_{\rm b}$ and $\tau_{\rm e}$ increase monotonically when decreasing $g_T$, the difference $\tau_{\rm b} - \tau_{\rm e}$, too, is independent of the initial condition for $g_T$. This implies that all trajectories give rise to essentially {\it the same} number of $e$-folds: $N \approx 0.9$. This number is far smaller than the 60 $e$-folds occurring in the ``usual'' inflationary scenarios, so that there is clearly no direct correspondence.  
\end{subsubsection}
\end{subsection}
\begin{subsection}{RG-improved IR cosmologies}
\label{sect5.4}
In this section we investigate the ``IR cosmologies'' related to the $k < k_T$ branch of the RG trajectories. For the numerical illustration we shall use the initial conditions \eqref{6.3} again, this time integrating the RG equations towards smaller $k$.
\begin{subsubsection}{The running of $G$ and $\Lambda$}
In fig.\ \ref{drei} we have plotted $G(k)$ and $\Lambda(k)$ for the three sample trajectories.
\begin{figure}[t]
\renewcommand{\baselinestretch}{1}
\epsfxsize=0.48\textwidth
\begin{center}
\leavevmode
\epsffile{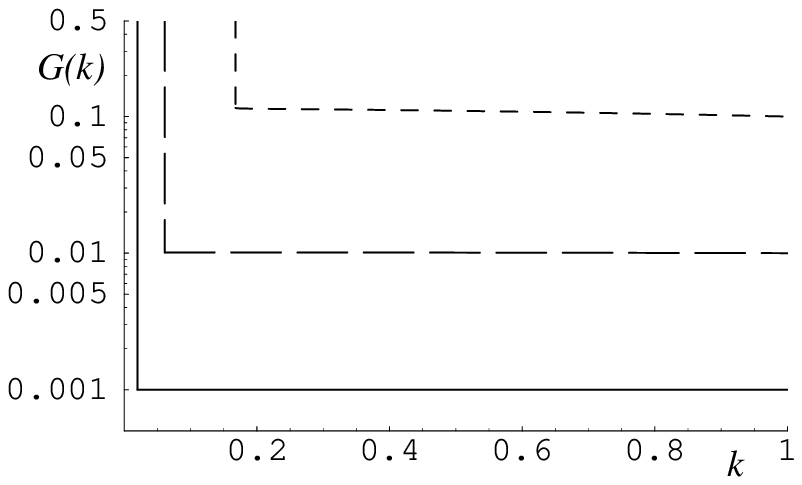} \,  
\epsfxsize=0.48\textwidth
\epsffile{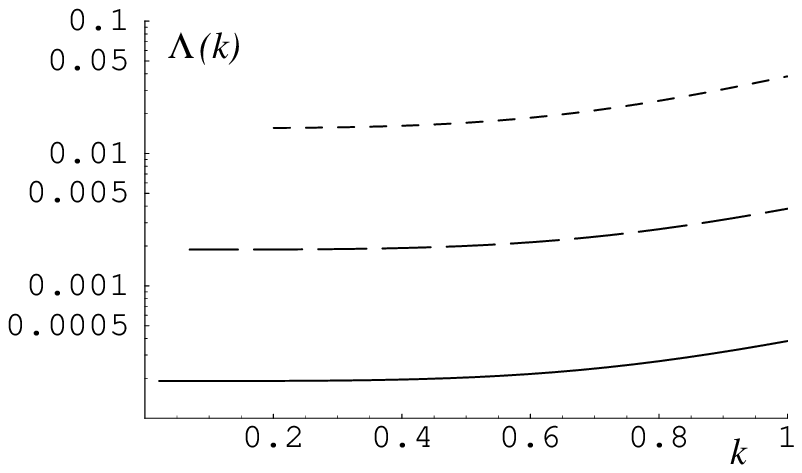}
\end{center}
\parbox[c]{\textwidth}{\caption{\label{drei}{\footnotesize The functions $G(k)$ and $\Lambda(k)$ arising from the initial conditions $g_T = 10^{-1}$ (short dashes), $g_T = 10^{-2}$ (long dashes) and $g_T = 10^{-3}$ (solid).}}}
\end{figure}
Going downward from $k=1$ (in $k_T$-units, as always) the cosmological constant has a weak scale dependence at first but then assumes a constant value over a wide range of $k$-values. For $G$, this plateau behavior has set in before the turning point even. These plateaus correspond to the classical GR regime where, by definition, $G, \Lambda \approx const$. At even smaller $k$, $G(k)$ starts increasing until it finally develops a vertical tangent at the termination scale $k_{\rm term}$. (For the real RG trajectory with $g_T \approx 10^{-60}$ the increase is less abrupt than it appears in fig.\ \ref{drei}.) For $k \gtrsim k_{\rm term}$, as long as $G(k)$ is not too different from its classical plateau value, the Einstein-Hilbert truncation should be reliable still. It is natural to ask, therefore, whether the IR increase of $G(k)$ leads to cosmological consequences, as long as we can trust the approximation. (Note that $\Lambda(k)$ remains regular at the termination point; $\lambda(k_{\rm term}) = 1/2$ entails $\Lambda(k_{\rm term}) = k^2_{\rm term}/2$.)
\end{subsubsection}
\begin{subsubsection}{The cosmological history}
Applying the method of Section 2 to the functions $G(k)$ and $\Lambda(k)$ of fig.\ \ref{drei} we obtain the cosmological solutions displayed in the figs.\ \ref{air} (scale factor), \ref{fuenf} (Hubble and deceleration parameter), \ref{vier} (time dependence of $G$ and $\Lambda$), and \ref{zwoelf} (scale-time relationship).
\begin{figure}[t]
\renewcommand{\baselinestretch}{1}
\begin{center}
\leavevmode
\epsfxsize=0.6\textwidth
\epsffile{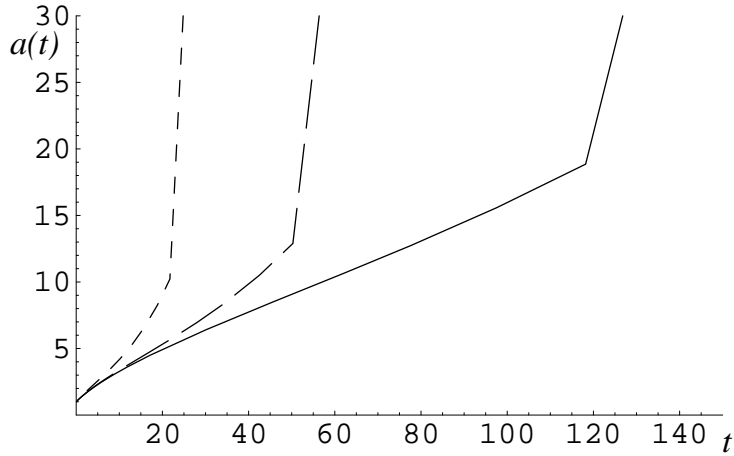} \,  
\end{center}
\parbox[c]{\textwidth}{\caption{\label{air}{\footnotesize IR cosmology: The scale factor $a(t)$  arising from $g_T = 10^{-1}$ (short dashes), $g_T = 10^{-2}$ (long dashes) and $g_T = 10^{-3}$ (solid line).}}}
\end{figure}
\begin{figure}[t]
\renewcommand{\baselinestretch}{1}
\epsfxsize=0.48\textwidth
\begin{center}
\leavevmode
\epsffile{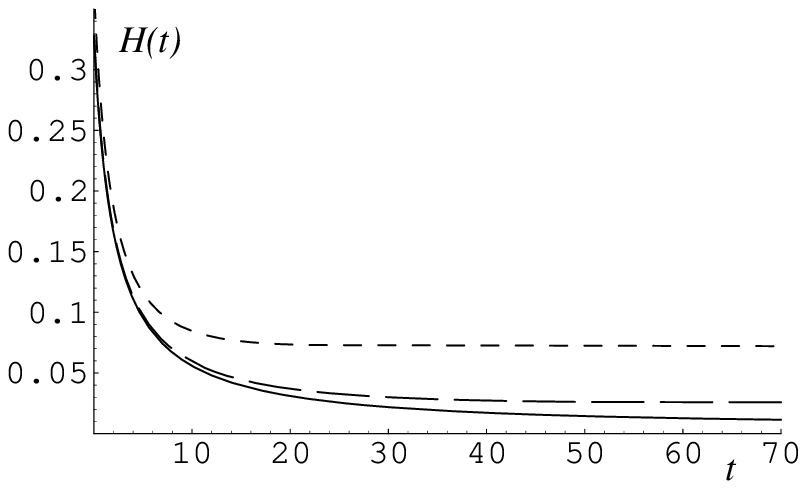}   
\epsfxsize=0.48\textwidth
\epsffile{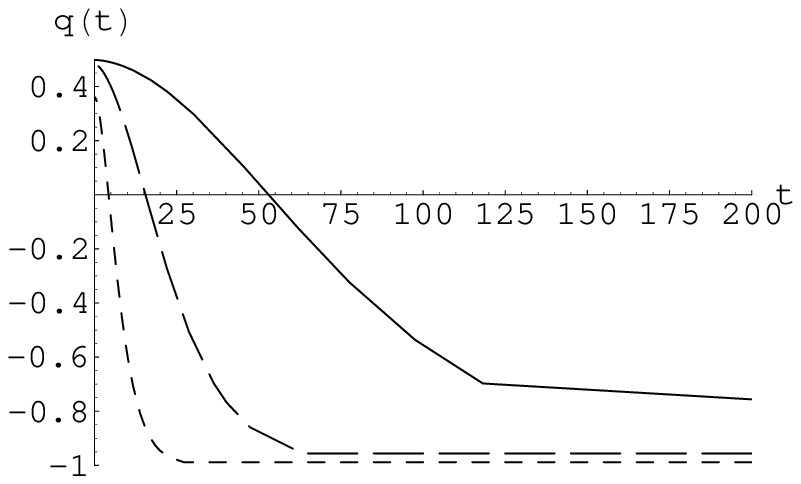}
\end{center}
\parbox[c]{\textwidth}{\caption{\label{fuenf}{\footnotesize IR cosmology: $H(t)$ and $q(t)$ for the cosmologies starting with  $g_T = 10^{-1}$ (short dashes), $g_T = 10^{-2}$ (long dashes) and $g_T = 10^{-3}$ (solid line). Decreasing $g_T$ increases the cosmological time-span passing until the Hubble parameter becomes constant and delays the transition to a universe dominated by the cosmological constant.}}}
\end{figure}

The gross features of all three of the typical cosmologies considered are as follows. Near $k \approx k_T$ the universe is matter dominated $(\Omega_M \approx 1, \Omega_\Lambda \approx 0)$ and has a deceleration parameter $q \approx 1/2$ therefore. According to fig.\ \ref{fuenf}, this value decreases towards $q = -1$ at asymptotically late times. This asymptotic regime is completely $\Lambda$-dominated: $\Omega_M \approx 0, \Omega_\Lambda \approx 1$. Decreasing $g_T$ the cosmological time passing during this transition from matter to vacuum dominance increases. The value $q \approx -0.55$, for instance, which roughly corresponds to the deceleration parameter observed today, is reached at increasingly late cosmological times as $g_T$ is becoming smaller.  The Hubble parameter approaches a constant value for $t \rightarrow \infty$, indicating that the space-time approaches de~Sitter space asymptotically. The kink in the $a(t)$-curves of fig.\ \ref{fuenf} marks the onset of the de~Sitter behavior.

The transition form matter to $\Lambda$-dominance occurs in classical cosmology as well. To what extent are the improved cosmologies modified by quantum gravity effects? Fig.\ \ref{vier} shows that $\Lambda(t)$ is decreasing as long as $k$ is not too far below $k_T$, while $G(t)$ is essentially constant.
\begin{figure}[t]
\renewcommand{\baselinestretch}{1}
\epsfxsize=0.48\textwidth
\begin{center}
\leavevmode
\epsffile{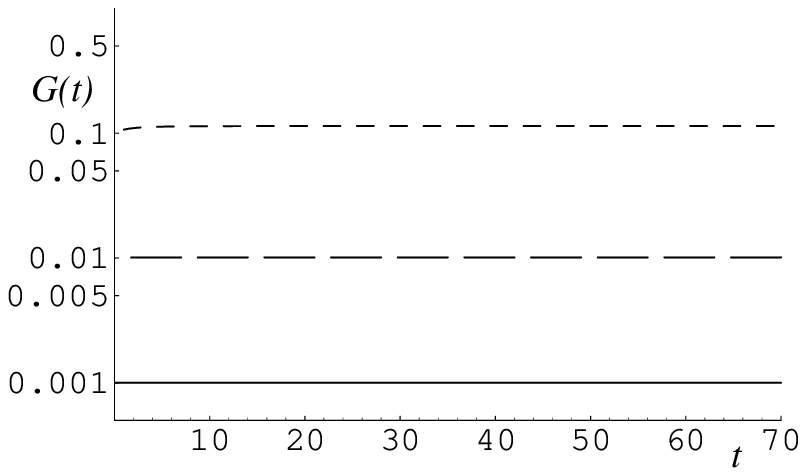}   
\epsfxsize=0.48\textwidth
\epsffile{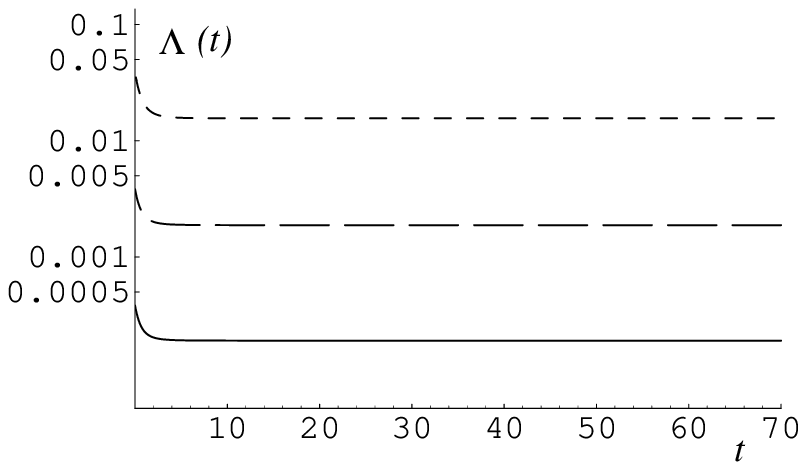}
\end{center}
\parbox[c]{\textwidth}{\caption{\label{vier}{\footnotesize IR cosmology: $G(t)$ and $\Lambda(t)$  arising from the initial conditions $g_T = 10^{-1}$ (short dashes), $g_T = 10^{-2}$ (long dashes) and $g_T = 10^{-3}$ (solid line). Note the logarithmic scaling of the vertical axis. Decreasing $g_T$ decreases the asymptotic value of $\Lambda(t)$ and $G(t)$ for $t \rightarrow \infty$.}}}
\end{figure}
For $t \gtrsim 5 k^{-1}_T$, say, both $\Lambda(t)$ and $G(t)$ attain constant values which remain unaltered for $t \rightarrow \infty$ (at least within the resolution of fig.\ \ref{vier}). In the next subsection we shall demonstrate that the asymptotic values of $\Lambda$ and $G$ are exactly those predicted by the flow equation linearized about the GFP provided $g_T$ is small enough. This implies that, for $g_T$ sufficiently small, i.e., for trajectories getting close to the GFP, there are no significant renormalization effects at late times.
\begin{figure}[t]
\renewcommand{\baselinestretch}{1}
\epsfxsize=0.6\textwidth
\begin{center}
\leavevmode
\epsffile{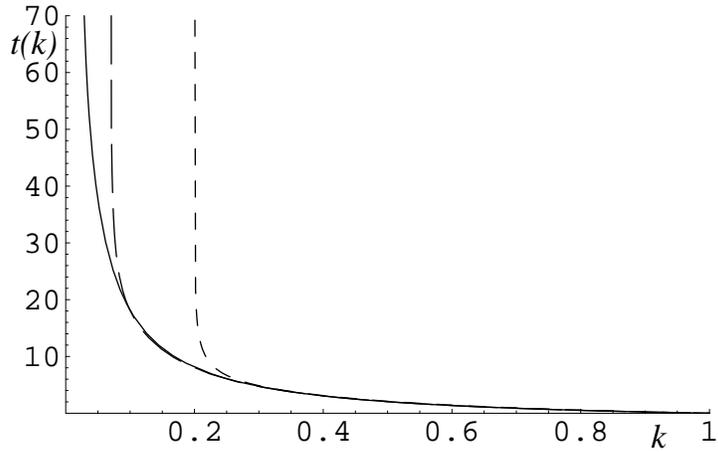} 
\end{center}
\parbox[c]{\textwidth}{\caption{\label{zwoelf}{\footnotesize The function $t(k)$ arising from the initial conditions $g_T = 10^{-1}$ (short dashes), $g_T = 10^{-2}$ (long dashes) and $g_T = 10^{-3}$ (solid). The cosmological time $t$ increases monotonically with decreasing $k$ and  diverges as the RG trajectory reaches the $\Omega$-line.}}}
\end{figure}
\end{subsubsection}
\begin{subsubsection}{Asymptotic vs.\ laboratory values and the cosmological constant problem}
Let us come back to the plots of $G(t)$ and $\Lambda(t)$ shown in fig.\ \ref{vier}.
This figure illustrates that decreasing $g_T$ decreases the asymptotic values of $\Lambda(t)$ and $G(t)$ for $t \rightarrow \infty$. Table \ref{one} summarizes these asymptotic values of $G$ and $\Lambda$ for various choices of $g_T$. This table also contains the corresponding ``laboratory'' values $G_{\rm lab}$ and $\Lambda_{\rm lab}$. Their interpretation is as follows \cite{h3}.
\begin{table}[t]
\begin{tabular*}{\textwidth}{@{\extracolsep{\fill}} c||c|c||c|c } 
$g_T$ & $G_{\rm asym}[k^{-2}_T]$ & $\Lambda_{\rm asym}[k^2_T]$ & $G_{\rm lab}[k^{-2}_T]$ & $\Lambda_{\rm lab}[k^2_T]$ \\ \hline 
$10^{-1}$ & $1.1 \times 10^{-1}$ & $1.5 \times 10^{-2}$ & $1.0 \times 10^{-1}$ & $1.91 \times 10^{-2}$ \\
$10^{-2}$ & $1.01 \times 10^{-2}$ & $1.9 \times 10^{-3}$ & $1.00 \times 10^{-2}$ & $1.91 \times 10^{-3}$ \\
$10^{-3}$ & $1.001 \times 10^{-3}$ & $1.9 \times 10^{-4}$ & $1.000 \times 10^{-3}$ & $1.91 \times 10^{-4}$ \\
$10^{-5}$ & $1.00001 \times 10^{-5}$ & $1.9 \times 10^{-6}$ & $1.000 \times 10^{-5}$ & $1.91 \times 10^{-6}$ \\
$10^{-7}$ & $1.0000001 \times 10^{-7}$ & $1.9 \times 10^{-8}$ & $1.000 \times 10^{-7}$ & $1.91 \times 10^{-8}$ \\
\end{tabular*}
\renewcommand{\baselinestretch}{1}
\parbox[c]{\textwidth}{\caption{\label{one}{\footnotesize The asymptotic and laboratory values of $G$ and $\Lambda$ in units of $k_T$ for various choices of $g_T$.}}}
\end{table}

Provided the trajectory gets sufficiently close to the GFP it contains a long classical regime.  (In fig.\ \ref{null} this ``GR-regime'' corresponds to the segment of the trajectory between the points P$_1$ and P$_2$.) In this regime the dimensionful quantities $G$ and $\Lambda$ are approximately $k$-independent. Linearizing about the GFP one finds for their values, expressed in terms of $k_T$,
\begin{align*}
\addtocounter{equation}{1} 
\tag{\theequation a} 
\label{5.1a} 
G_{\rm lab} & = g_T / k_T^2 \, ,\\
\tag{\theequation b}
\label{5.1b} 
\Lambda_{\rm lab} & = \frac{1}{2} \lambda_T k^2_T = (\varphi_2 / 4 \pi) g_T k^2_T \, .
\end{align*} 
In table \ref{one} we compare the asymptotic values of $G$ and $\Lambda$ to their ``lab'' values computed from (\ref{5.1a},b). We find that for $g_T$ of the order $10^{-2}$ or smaller the plateau values at asymptotically late times virtually coincide with those measured in a ``laboratory'' in which classical GR is known to apply. At $g_T = 10^{-1}$, instead, $G_{\rm asym}$ is slightly larger than $G_{\rm lab}$ while $\Lambda_{\rm asym}$ is slightly smaller than $\Lambda_{\rm lab}$.

The interpretation of this result is as follows. For $g_T$ sufficiently small, the universe does not experience significant IR renormalization effects; in the limit $t \rightarrow  \infty$ it basically keeps its values of $G$ and $\Lambda$ from the GR regime. Only for comparatively large values of $g_T$ we find deviations between the GR-formulas (\ref{5.1a},b) and the numerically computed asymptotic values. Those deviations occur because the RG trajectory does not get sufficiently close to the GFP to give rise to a long classical regime where (\ref{5.1a},b) would apply. 

The conventional Planck mass is defined in terms of Newton's constant measured in a classical ``laboratory'': $m_{\rm Pl} \equiv 1/\sqrt{G_{\rm lab}}$. This definition, together with \eqref{5.1a} leads to the following important relation between $m_{\rm Pl}$ and $k_T$:
\be\label{5.2}
k_T^2 = g_T \, m_{\rm Pl}^2 \, .
\ee
Obviously $g_T \ll 1$ leads to a large hierarchy $k_T / m_{\rm Pl} \ll 1$. In table \ref{two} we use \eqref{5.2} in order to express the asymptotic and laboratory values of $G$ and $\Lambda$ in terms of the more familiar Planckian units. 

As for the cosmological constant problem, the crucial point to be noted is that $\Lambda_{\rm lab} \approx \Lambda_{\rm asym}$ is suppressed by a factor $\propto g_T^2$ relative to its ``natural'' value $m_{\rm Pl}^2$. As was discussed in detail in ref.\ \cite{h3}, fine-tuning the RG-trajectory in such a way that it spends a long RG time near the GFP by picking an ``unnaturally'' small value of $g_T$ leads to a long (in $k$-units) classical regime on the trajectory. Once this is achieved, the solution of the cosmological constant problem is for free:
\be\label{5.3}
\Lambda_{\rm lab} \, / \, m_{\rm Pl}^2 = (\varphi_2/4 \pi) \, g_T^2 \ll 1 \, .
\ee
\begin{table}[t]
\begin{tabular*}{\textwidth}{@{\extracolsep{\fill}} c||c|c||c|c } 
$g_T$ & $G_{\rm asym}[m_{\rm Pl}^{-2}]$ & $\Lambda_{\rm asym}[m_{\rm Pl}^2]$ & $G_{\rm lab}[m_{\rm Pl}^{-2}]$ & $\Lambda_{\rm lab}[m_{\rm Pl}^2]$ \\ \hline 
$10^{-1}$ & $1.1 \times 10^{-1}$ & $1.5 \times 10^{-3}$ & $1$ & $1.91 \times 10^{-3}$ \\
$10^{-2}$ & $1.01 \times 10^{-2}$ & $1.9 \times 10^{-5}$ & $1$ & $1.91 \times 10^{-5}$ \\
$10^{-3}$ & $1.001 \times 10^{-3}$ & $1.9 \times 10^{-7}$ & $1$ & $1.91 \times 10^{-7}$ \\
$10^{-5}$ & $1.00001 \times 10^{-5}$ & $1.9 \times 10^{-11}$ & $1$ & $1.91 \times 10^{-11}$ \\
$10^{-7}$ & $1.0000001 \times 10^{-3}$ & $1.9 \times 10^{-15}$ & $1$ & $1.91 \times 10^{-15}$ 
\end{tabular*}
\renewcommand{\baselinestretch}{1}
\parbox[c]{\textwidth}{\caption{\label{two}{\footnotesize The asymptotic and laboratory values of $G$ and $\Lambda$ in Planck units for various choices of $g_T$. Decreasing $g_T$ decreases both the asymptotic and the laboratory value of the cosmological constant.}}}
\end{table}
The numbers in table \ref{two} illustrate this important phenomenon. The smaller is $g_T$, the closer the trajectory approaches the GFP, the smaller is the cosmological constant in the classical GR laboratory further downstream on the trajectory.

The relative importance of the cosmological constant can be summarized as follows. In the very early universe, in the NGFP regime, one has $\Omega_M = \Omega_\Lambda = 1/2$ showing that the vacuum and matter energy densities drive the expansion on an exactly equal footing. Near the GFP, in particular in the classical GR regime, one automatically is lead to $\Omega_\Lambda \approx 0$ and $\Omega_M \approx 1$, a conventional purely matter dominated (flat) FRW cosmology. At infinitely late times $\Lambda$ takes over again and with $\Omega_\Lambda \approx 1$ and $\Omega_M \approx 0$ a de~Sitter like behavior sets in. 
\end{subsubsection}
\begin{subsubsection}{$\Omega$-line mechanism and dynamical cutoff identification}
In fig.\ \ref{drei} we saw that $G(k)$ and $\Lambda(k)$ show a strong IR running for small $k$. It therefore comes as a surprise perhaps that $G(t)$ and $\Lambda(t)$, according to fig.\ \ref{vier}, show no sign of such IR renormalizations for $t \rightarrow \infty$; according to the previous subsection, the late cosmology is essentially classical.

This apparent contradiction is resolved in fig.\ \ref{zwoelf} which displays the relationship between the scale $k$ and the cosmological time $t$. As expected, $t = t(k)$ is a monotonically decreasing function of $k$. However, an infinitely old universe (``$t = \infty$'') does not correspond to $k = 0$, but rather to a non-zero asymptotic scale
\be\label{5.50}
k_{\rm asym} \equiv k(t \rightarrow \infty) > 0 \, .
\ee 
The function $t = t(k)$ has a singularity at this scale: $t(k \searrow k_{\rm asym}) \rightarrow \infty$. Stated differently, the inverse function $k = k(t)$ is bounded below by $k_{\rm asym}$ and, even for arbitrarily late times, does not reach arbitrarily small scales: $k(t) > k_{\rm asym}$ for all $t$.

The absence of visible IR effects in $G(t) \equiv G(k = k(t))$ and $\Lambda(t) \equiv \Lambda(k = k(t))$ is explained by the fact that, for the examples considered, $k_{\rm asym}$ is much larger that the scale $k_{\rm term}$ near which $G(k)$ and $\Lambda(k)$ get renormalized strongly. This is precisely the $\Omega$-line mechanism we discussed already earlier: for $t \rightarrow \infty$ the dimensionless $g(k(t))$ and $\lambda(k(t))$ approach a point on the $\Omega$-line. The endpoint of the cosmology, ``$a = \infty$'' or $\rho = 0$, corresponds to this point on the $g$-$\lambda$-plane; it is still far away from the $|\eta_N| = \infty$-line close to which the IR renormalizations would become strong. Therefore, once the universes have entered the GR regime at about $t \approx 5 k_T^{-1}$, say, they remain classical for all later times. In this sense the $\Omega$-line separates the cosmologically accessible parts of theory space from those with strong IR running.  

The asymptotic de~Sitter phase of the universe corresponds to the approximately time independent scale $k(t) \approx const = k_{\rm asym} \gg k_{\rm term}$.

Let us look more closely at the cutoff identification $k = k(t)$ which was generated dynamically by our system of equations. In a slightly different presentation it is displayed as the $k < k_T$-branch in fig.\ \ref{acht}. In this figure $(t(k) k)^{-1}$ and $H(t(k))/k$ are plotted vs.\ $\ln(k/k_T)$. If the actual (obviously rather complicated) cutoff identification was close to $k \propto 1/t$ and $k \propto H(t)$, respectively, those plots would show a constant function for all $k$.

It is clear, and also confirmed by the plots, that in the IR the real $k(t)$ is quite different from $k \propto 1/t$, the reason being that the real $k(t)$ is constant, equal to $k_{\rm asym}$, while $1/t$ decreases monotonically for $t \rightarrow \infty$.

On the other hand, $H(t(k))/k$ is seen to be approximately $k$-independent within a factor of less than about 2. Thus, at least at a qualitative or ``semi-quantitative'' level, the actual cutoff identification can be approximated by $k \propto H$.\footnote{As an ansatz, $k \propto H$ has also been employed in refs. \cite{sola} in a different context.} This identification would indeed associate a {\it constant} scale $k_{\rm asym}$ to the de~Sitter final state, proportional to the time independent $H_{\rm asym} = \sqrt{\Lambda_{\rm asym}/3}$.

From these observations it is clear how the cosmologies are to be interpreted. Since approximately $k \propto H$ the length scale characterizing the averaging volume is the Hubble radius $\ell_H \equiv H^{-1}$. The cosmological parameters computed correspond to averages over the volume $\ell^4_H$. The scale $\ell_H$ characterizes the radius of curvature of the four-dimensional space-time, the size of the ``Einstein elevator''. The asymptotic de~Sitter phase makes it particularly clear that the temporal proper distance to the Big Bang, $t$, would not lend itself as a coarse graining scale within the improved equations approach. For the UV cosmology, in particular near the NGFP \cite{cosmo1}, both identifications are equivalent, though.
\end{subsubsection}
\end{subsection}
\begin{subsection}{The RG-trajectory realized by Nature}
In ref.\ \cite{h3} the observational (supernova, CMBR, etc.) data were interpreted under the assumption that the gravitational RG trajectory which Nature realizes has the qualitative features of a Type IIIa trajectory of QEG. It was found that the turning point of this trajectory is extremely close to the GFP, and that it is passed at a scale very far below the Planck scale:
\be\label{5.60}
g_T = \cO(10^{-60}) \, , \quad \lambda_T = \cO(10^{-60}) \, , \quad k_T = \cO(10^{-30} m_{\rm Pl}) \, .
\ee
These order of magnitude estimates do not depend on whether the late universe is, or is not affected by IR renormalization effects. The analysis shows that even if there was no $\Omega$-mechanism the data would not afford a ratio $G_{\rm cosmological}/G_{\rm lab}$ larger than about one order of magnitude \cite{h3}.

The universe passes the turning point a fraction of a second after the Big Bang, at a Hubble radius of the order $\ell_H \approx k_T^{-1} \approx 10^{30} l_{\rm Pl} \approx 10^{-3}$ cm. After that $g(k(t))$ decreases from $10^{-60}$ to its present value $g_{\rm today}$ of about $10^{-120}$.

Since $g(k) \ll 1$ during most epochs of the cosmological evolution, in particular in the late (IR) universe, we may use the $g \rightarrow 0$ limit of the corresponding beta-functions there. This allows for a simple determination of the point on the $g$-$\lambda$-plane where the universe resides today. Assuming we know $\Omega_{\Lambda 0} \equiv \Omega_{\Lambda}(t_{\rm today})$ we can solve the second equation of \eqref{4.y} for the present value of $\lambda$, $\lambda_{\rm today}$:
\be\label{5.61}
1- y(0, \lambda_{\rm today}) = \Omega_{\Lambda 0}^{-1} \, .
\ee
Here we approximated $g_{\rm today} \approx 0$. Note that the precise value of $\lambda_{\rm today}$ is scheme dependent.

For a rough estimate of $\lambda_{\rm today}$ we may assume that the classical FRW cosmology is essentially correct at late times. Analyzing the observational data within this framework yields $\Omega_{\Lambda 0} \approx 0.7$. For this value, and the $y$-function corresponding to the sharp cutoff, we find
\be\label{4.11}
\lambda_{\rm today} \approx 0.320 \, .
\ee
Looking at fig.\ \ref{eins} confirms that the point $(g_{\rm today}, \lambda_{\rm today}) \approx (10^{-120}, 0.32)$ indeed lies to the left of the $\Omega$-line for the sharp cutoff, as it should be.
\begin{subsubsection}{The age of the RG-universe}
As a check of our beta-function formalism, and in order to confirm the conclusions of subsection \ref{sect5.4}, we next derive a relationship between the age of the universe, $t_{\rm today}$,  the cosmological constant $\Lambda_{\rm lab} \approx \Lambda_{\rm asym}$, and the measured $\Omega_{\Lambda 0}$.

First we use the value of $\Omega_{\Lambda 0}$ in order to solve \eqref{5.61} for $\lambda_{\rm today}(\Omega_{\Lambda 0})$. Next we rewrite (the reciprocal) eq.\ \eqref{1.2} in terms of the dimensionless couplings $g = k^2 G$, $\lambda = \Lambda / k^2$ and expand its RHS up to the leading order in $g$:
\be\label{4.12}
\begin{split}
\frac{dt}{dk} = & \frac{\sqrt{3}\,k^{-2}}{  \sqrt{\lambda} \, (1 - y_0)^{1/2}} 
\left[ 
\frac{(\tilde{B}_1^{\rm sc})^{\prime}}{B_1^{\rm sc}} 
- \frac{\pi \lambda \, (\tilde{B}_1^{\rm sc})^{\prime} 
- \left( 10 \ln(1-2 \lambda) - 4 \zeta(3) + \frac{20 \lambda}{1-2 \lambda} \right)}{\pi \lambda \, B_1^{\rm sc} - \left(  5 \ln(1-2 \lambda) - 2 \zeta(3)  \right)  } 
\right] \\
& + \cO(g) \, .
\end{split}
\ee
Here
\be\label{4.13}
(\tilde{B}_1^{\rm sc})^{\prime} = \frac{4}{3 \pi} \, \lambda \left[ - \, \frac{5}{1 - 2 \lambda}  + \frac{18}{(1 - 2 \lambda)^2}\right] \, .
\ee
This expression can be converted to a differential equation for $dt/d \lambda$ by exploiting that in the GR regime we have the relation $\lambda = \Lambda_{\rm lab} / k^2$. Its use is legitimate here since the age of the universe for trajectories with sufficient ``squeezing'' is dominated by their GR regime. Substituting it into \eqref{4.13} we obtain the desired general relationship:
\be\label{4.14}
t_{\rm today} =  \frac{1}{\sqrt{\Lambda_{\rm lab}}} \, I(\lambda_{\rm today}) \, .
\ee
Here $I(\lambda_{\rm today})$ is given by the dimensionless integral
\be\label{4.15}
\begin{split}
I(\lambda_{\rm today}) \equiv & - \, \int_0^{\lambda_{\rm today}} \,   d \lambda \, \frac{\sqrt{3}}{2 \lambda (1-y_0)^{1/2}} \\
&  \times
\left[ 
\frac{(\tilde{B}_1^{\rm sc})^{\prime}}{B_1^{\rm sc}} 
- \frac{\pi \lambda \, (\tilde{B}_1^{\rm sc})^{\prime} 
- \left( 10 \ln(1-2 \lambda) - 4 \zeta(3) + \frac{20 \lambda}{1-2 \lambda} \right)}{\pi \lambda \, B_1^{\rm sc} - \left(  5 \ln(1-2 \lambda) - 2 \zeta(3)  \right)  } 
\right] \, .
\end{split}
\ee

As an example, we evaluate this integral numerically for the $\lambda_{\rm today}$ given in eq. \eqref{4.11}. We find
\be\label{4.16}
I(\lambda_{\rm today} = 0.32) =  4.19
\ee
Using this value in \eqref{4.15}, along with $\sqrt{\Lambda_{\rm lab}} \approx \sqrt{\Lambda_{\rm asym}} \approx 10^{-60} m_{\rm Pl}$ we obtain the age $t_{\rm today} = 4.19 \, \times \, 10^{60} \, t_{\rm Pl}$. As expected, this is essentially the same age one obtains from the classical FRW equations if one uses the same input data.
\end{subsubsection}
\begin{subsubsection}{The deceleration parameter}
We now calculate $q(t)$ along the trajectory realized in Nature below the turning point. As in the previous subsection we will work in the approximation $g \searrow 0$. We parameterize the trajectory by its $\lambda$-values $\lambda \in [\lambda_T , \lambda_{\Omega-{\rm line}}]$, i.e., $k$ is thought of as a function of $\lambda$, with the inverse $\lambda = \lambda(k)$. In this parameterization we can use \eqref{4.14} to obtain $t(\lambda(k)) = 1/\sqrt{\Lambda_{\rm lab}} I(\lambda(k))$, the cosmological time passing during the evolution along the trajectory, while the corresponding deceleration parameter $q(\lambda(k))$ is obtained from eq.\ \eqref{9.5c}.
\begin{figure}[t]
\renewcommand{\baselinestretch}{1}
\epsfxsize=0.6\textwidth
\begin{center}
\leavevmode
\epsffile{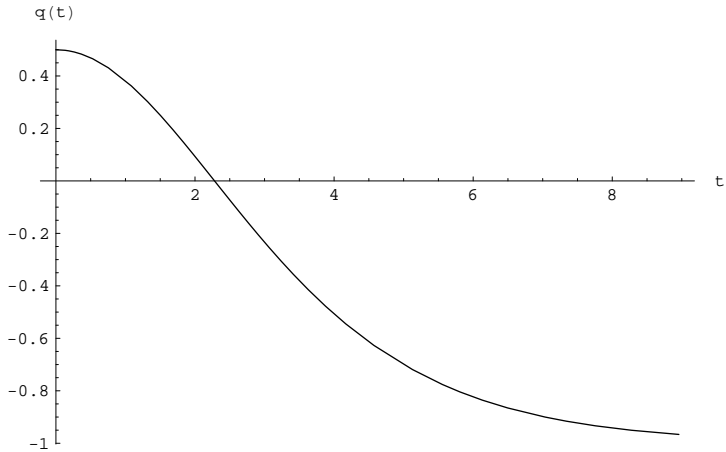}
\end{center}
\parbox[c]{\textwidth}{\caption{\label{elf}{\footnotesize The deceleration parameter $q(t)$ along the RG-trajectory realized in Nature. The time along the horizontal axis is given in units $10^{60} t_{\rm Pl}$. At present, $t \approx 4.2$, corresponding to $q \approx -0.55$.}}} 
\end{figure}
The function $q(t)$ along the trajectory can then be obtained as the parametric curve $\lambda \mapsto \big( t(\lambda), q(\lambda) \big)$, $\lambda \in [\lambda_T , \lambda_{\Omega-{\rm line}}]$. 

The resulting $q(t)$ is shown in fig.\ \ref{elf}. Note that the time $t$ displayed on the horizontal axis is in units $10^{60} t_{\rm Pl}$, indicating that the ``crossover'' from $q(0) = 1/2$ to $q(t \rightarrow \infty) = -1$ indeed occurs at cosmological time spans of the order of the age of our present universe. Indeed, the present universe corresponds to the age $t_{\rm today} \approx 4.2 \times 10^{60} t_{\rm Pl}$ and the deceleration parameter $q_{\rm today} \approx -0.55$.
\end{subsubsection}
\end{subsection}
\end{section}

\section{Summary and Discussion}
In this paper we presented a coherent renormalization group based framework which allows for the inclusion of potential quantum gravity effects during the entire cosmological history, from the epoch after the initial singularity to the scales of the present universe. We demonstrated that the strategy of RG improving the field equations is physically viable during all stages of the cosmological evolution. While from a mathematical point of view it is always possible to fix the cutoff scale in a way which renders the modified Einstein equations consistent, the result of the dynamical determination of $k(t)$ is not a priori guaranteed to lead to a useful interpretation of the coarse grained cosmology. However, we found that the dynamical cutoff identification corresponds to a time dependent averaging scale $\ell(t)$ which, at any time, is approximately equal to the Hubble radius $\ell_H(t)$. Therefore we may conclude that the ``microscope'' whose observations are described by the RG improved cosmology has exactly the right ``resolving power'' to see the large scale structure typical of cosmology.

One of the most interesting features of the early cosmology was the discovery of a kind of ``oscillatory inflation''. Running time backwards, the initial singularity is approached by an infinite sequence of increasingly short time intervals during which the universe accelerates and decelerates, respectively. This behavior is due to the fact that the non-Gaussian fixed point has complex critical exponents. The oscillatory approach of the Big Bang is reminiscent of the classical Belinskii-Khalatnikov-Lifshitz scenario \cite{bkl} but contrary to the latter the quantum effect found here occurs even in isotropic universes. 

The most interesting aspect of the improved cosmologies at late times is the ``$\Omega$-mechanism'' which we investigated in detail. At small scales the RG trajectories of the relevant type leave the domain of validity of the Einstein-Hilbert truncation, and it has been argued that in the IR there could be a regime where the RG effects become strong again. Remarkably, even if this regime exists, the cosmologies we found would never get affected by it. Even at arbitrarily late times they remain essentially classical since the underlying RG trajectory is never probed beyond a certain ``$\Omega$-point'' at which $\eta_N$ is very small still.

At this point we must ask whether this shielding of this $\eta_N$-divergence is a reliable prediction. Does the universe as a whole really never enter the regime where $|\eta_N|$ is large? Our argument was based upon two crucial facts: ({\it i}) The consistency condition for the RG improved Einstein's equation has the simple form \eqref{2.1c} which leads directly to the relationship $\rho \propto \Lp(k) / \Gp(k)$ of eq.\ \eqref{2.4b}; ({\it ii}) The RG flow is such that the zero of $\Lp(k)$ occurs at a {\it higher} value of $k$ than the divergence of $\Gp(k)$. As a consequence of ({\it ii}), $\rho$ vanishes and $a \propto \rho^{-1/(3 + 3w)}$ reaches infinity before $|\eta_N|$ has become large.

As for the validity of ({\it ii}), even though we analyzed the Einstein-Hilbert truncation only, this prediction was perfectly stable under a change of the cutoff scheme. Because of this robustness, it is likely to be actually correct and not a truncation artifact. 

The much more subtle issue is ({\it i}) which is not related to the RG flow per se but to the improvement procedure. It is important to note that, at least in this particular form, the zero of $\Lp(k)$ implies an infinite scale factor only in the approach of improving the \emph{field equations}. It would not happen if one improves the solutions of the classical equations by replacing $G \mapsto G(t)$, $\Lambda \mapsto \Lambda(t)$ there, or if one improves the action functional. The latter approach has been investigated in detail in ref.\ \cite{h1}. Performing the substitution under the space-time integral of the Einstein-Hilbert action one obtains a different consistency condition:\footnote{This form of the consistency condition corresponds to the choice $\Theta_{\mu \nu} = 0$, see ref.\ \cite{h1}.} 
\be\label{8.1}
\Ld + 8 \pi \, \rho \, \Gd + 3 \, H \, \left( \frac{\Gd}{G} \right)^2 + 3 \, \left( \frac{\Gd}{G}\right) \, \left( \frac{\add}{a} \right) = 0 \, .
\ee
We observe that, in this framework, $\Ld \propto \Lp \rightarrow 0$ by no means implies that $\rho \rightarrow 0$ and $a \rightarrow \infty$. The improved equation and action approaches are (almost) equivalent in the UV, but can lead to different predictions in the IR. There they effectively represent certain non-local terms of the ordinary effective action $\Gamma_{k=0}$ generated during the RG running but not taken into account explicitly \cite{h1}. The additional terms in \eqref{8.1} which are not present in \eqref{2.1c} are due to the fact that in the improved action approach the scalar fields $G$ and $\Lambda$ carry energy and momentum which is neglected in the improved equation approach. 

For this reason it cannot be excluded for the time being that the shielding of the $\eta_N$-divergence is an artifact of the improvement scheme since, at least via the mechanism discussed above, the shielding at $\Lambda^\prime = 0$ occurs only if one neglects the energy and momentum carried by the scalars $G$ and $\Lambda$.  Clearly more work is needed in order to settle this issue. It should be emphasized, though, that whatever is the final answer about RG effects in late cosmology it bears no simple relationship to the conjectured modifications of gravity at galactic scales. The corresponding cutoff identifications mimicking the essential IR-relevant invariants in both cases are extremely difficult to guess beforehand, and in particular in a Lorentzian setting where the actual cutoff is (a generalization of) ``virtuality'' rather than ``inverse distance'' one certainly cannot expect that the visibility of the quantum effects increases with distance in a naive way. For this reason the results of the present paper have no direct relevance to the scenario of modified galactic dynamics in \cite{h2,h3}.
\bigskip

\noindent
{\large\bfseries Acknowledgements} \smallskip

\noindent
We would like to thank A.\ Bonanno, J.\ Moffat, and T.\ Prokopec for helpful discussions.

\begin{appendix}
\section{Threshold functions for various cutoff schemes}
\label{apA}
In this appendix we collect the relevant properties of the threshold functions $\Phi$ and $\tilde{\Phi}$ for the cutoff schemes used in the main part of the paper. We employ different cutoff schemes in order to gain insight about whether properties of our solutions are independent of the scheme used to suppress the IR-modes in the path integral when deriving the beta-functions (\ref{4.2a},b). This cutoff-scheme dependence is contained in the threshold functions
\be\label{A.1}
\begin{split}
\Phi^p_n(w) \equiv & \frac{1}{\Gamma(n)} \, \int^{\infty}_{0} dz \, z^{n-1} \, \frac{R^{(0)}(z) - z \, R^{(0)\prime}(z)}{\left[ z + R^{(0)}(z) + w\right]^p} \, , \\
\tilde{\Phi}^p_n(w) \equiv & \frac{1}{\Gamma(n)} \, \int^{\infty}_{0} dz \, z^{n-1} \, \frac{R^{(0)}(z) }{\left[ z + R^{(0)}(z) + w\right]^p} \, ,
\end{split}
\ee
(defined for $p = 1,2,3, \cdots$ and $n > 0$) through the choice of the dimensionless cutoff function $R^{(0)}(z)$. Concretely, we employed three different type A cutoffs \cite{oliver1, frank1}, the exponential cutoff, the optimized cutoff \cite{optcutoff}, and the sharp cutoff \cite{frank1}. The latter two have the advantage that the integrals appearing in \eqref{A.1} can be evaluated analytically. This provides a considerable simplification when solving the RG-equations as for these cases the RG-flow of $G(k), \Lambda(k)$ is governed by simple first order differential equations.

\begin{subsection}{The optimized cutoff}
For the optimized cutoff the dimensionless cutoff function $R^{(0)}(z)$ is given by \cite{optcutoff}
\be\label{A.2}
R^{(0)}(z)^{\rm opt} \equiv (1 - z) \, \Theta(1 - z) \, .
\ee
Substituting this expression into \eqref{A.1} and carrying out the integrals we obtain
\be\label{A.3}
\Phi^p_n(w)^{\rm opt} = \frac{1}{\Gamma(n+1)} \, \frac{1}{(1 + w)^p} \, , \quad 
\tilde{\Phi}^p_n(w)^{\rm opt} = \frac{1}{\Gamma(n+2)} \, \frac{1}{(1 + w)^p} \, .
\ee
These threshold functions are related through the relations
\be\label{A.4}
\begin{split}
\Phi^{p+1}_n(w)^{\rm opt} = & - \, \frac{1}{p} \, \frac{\partial}{\partial w}  \Phi^p_n(w)^{\rm opt} \, , \quad
 \Phi^p_{n+1}(w)^{\rm opt} =  \frac{1}{n+1} \Phi^p_n(w)^{\rm opt} \, , \\
\tilde{\Phi}^p_n(w)^{\rm opt} = & \frac{1}{n+1} \Phi^p_n(w)^{\rm opt} \, .
\end{split}
\ee
\end{subsection}
\begin{subsection}{The exponential cutoff}
The exponential cutoff is given by the following one-parameter family of cutoff functions \cite{souma,oliver1,frank1}
\be\label{A.5}
R^{(0)}(z;s)^{\rm Exp} \equiv \frac{s \, z}{\exp(s z) - 1} \; , \quad s > 0 \, , 
\ee
where $s$ is a shape parameter. In this case the integrals \eqref{A.1} cannot be evaluated analytically. Therefore we have to rely on a numerical integration when solving the RG equations. In order to have a reasonable convergence in the threshold functions we choose the shape parameter $s=2$.
\end{subsection}
\begin{subsection}{The sharp cutoff}
The sharp cutoff is defined via the cutoff function
\be\label{A.6}
R_k(p^2)^{\rm sc} \equiv k^2 \, R^{(0)}(z)^{\rm sc} \equiv \hat{R} \, \Theta(1 - z) \, ,
\ee
where the limit $\hat{R} \rightarrow \infty$ is taken after substituting this function into the integrals defining the threshold functions.  The resulting threshold functions have been determined in \cite{frank1}:
\be\label{A.7}
\begin{split}
\Phi^1_n(w)^{\rm sc} = & - \, \frac{1}{\Gamma(n)} \, \ln(1+w) + \varphi_n \, , \quad \mbox{for} \; p = 1 \, , \\
\Phi^p_n(w)^{\rm sc}  = & \frac{1}{\Gamma(n)} \, \frac{1}{p-1} \, \frac{1}{(1 + w)^{p-1}} \, , \quad \mbox{for} \; p > 1 \, , \\
\tilde{\Phi}^1_n(w)^{\rm sc}  = &  \frac{1}{\Gamma(n+1)} \, , \quad \mbox{for} \; p = 1 \, , \\
\tilde{\Phi}^1_n(w)^{\rm sc}  = &  0 \, , \quad \mbox{for} \; p > 1 \, . 
\end{split}
\ee
The $\varphi_n$'s are a priori undetermined constants of integration. We choose them as \cite{frank1}
\be\label{A.8}
\varphi_n \equiv \Phi^1_n(0)^{{\rm Exp} (s=1)} = n \, \zeta(n+1) \, ,
\ee
where $\zeta(n)$ denotes the Riemann $\zeta$-function.
\end{subsection}
\section{$H$ as a function of the RG scale}
\label{AppB}
In this appendix we rewrite the Hubble parameter as a function of $k, G(k)$, and $\Lambda(k)$. Our starting point is eq.\ \eqref{2.1a} which (for $K = 0$) yields 
\be
H \equiv \frac{\ad}{a}  =  \pm \, \frac{1}{\sqrt{3}} \, \left[  \Lambda + 8 \pi G \rho \right]^{1/2} \, ,
\ee
with the plus (minus) sign corresponding to an expanding (contracting) universe.
We then substitute eq.\ \eqref{2.4b} to obtain $H$ in terms of $G, \Lambda$ and their $k$-derivatives 
\be\label{B.9}
H = \pm \left[ \frac{1}{3 \Gp} \, \left( \Gp \, \Lambda - G \, \Lp \right) \right]^{1/2} \, .
\ee
The $k$ derivatives are conveniently expressed through the modified beta-functions
\be\label{B.1}
\bt_{G} \equiv \frac{1}{k} \, \beta_{G} \; , \quad \bt_{\Lambda} \equiv \frac{1}{k} \, \beta_{\Lambda} \, .
\ee
With these new functions the derivatives $G^{\prime}$, $\Lambda^{\prime}$ are simply
\be\label{B.2}
\Gp = \bt_{G}(k, G, \Lambda) \; , \qquad \Lp = \bt_{\Lambda}(k, G, \Lambda) \, . 
\ee
Substituting the sharp cutoff \eqref{A.7} into these equations we find
\be\label{B.3}
\begin{split}
\bt_{G} & =  \frac{k \, G^2 \, B_1^{\rm sc}}{1 - B_2^{\rm sc} \, k^2 \, G } \, \\ 
\bt_{\Lambda} & =  \frac{k \, G \, \Lambda \, B_1^{\rm sc}}{1 - B_2^{\rm sc} \, k^2 \, G }
+ \frac{1}{\pi} \, k^3 \, G \left[ -5 \ln \left( 1 - \frac{2 \Lambda}{k^2} \right) + 2 \zeta(3) - \frac{5}{4} \, \eta^{\rm sc} \right] \, .
\end{split}
\ee
where
\be\label{B.4}
\begin{split}
B_1^{\rm sc} & =  \frac{1}{3 \pi} \, \left[ -5 \ln \left( 1 - \frac{2 \Lambda}{k^2} \right) + \zeta(2) - \frac{18}{1 - \frac{2 \Lambda}{k^2}} - 6 \right] \, , \\
B_2^{\rm sc} & =  - \, \frac{5}{6 \pi} \, , \\
\eta^{\rm sc} & =  \frac{k^2 \, G \, B_1^{\rm sc}}{1 - B_2^{\rm sc} \, k^2 \, G } \, .
\end{split}
\ee
Eq.\ \eqref{B.9} can then be evaluated along any given RG-trajectory $k \mapsto (G(k), \Lambda(k))$. Thus, in terms of \eqref{B.3} with \eqref{B.4}, the Hubble parameter for an expanding universe is given by
\be\label{B.5}
H(k,G,\Lambda) = \left[ \frac{\Lambda \, \bt_G(k,G,\Lambda) - G \bt_{\Lambda}(k,G,\Lambda) }{3 \, \bt_G(k,G,\Lambda)} \right]^{1/2} \, .
\ee 
\end{appendix}
\end{document}